\documentclass[preprint,aps]{revtex4}
\usepackage{graphicx}
\usepackage{booktabs}
\usepackage[usenames, dvipsnames]{color}
\usepackage{multirow}
\usepackage{setspace}
\usepackage{rotating}
\usepackage[ pdftex, plainpages = false, pdfpagelabels, 
                 pdfpagelayout = useoutlines,
                 bookmarks,
                 bookmarksopen = true,
                 bookmarksnumbered = true,
                 breaklinks = true,
                 linktocpage=all,
                 pagebackref=false,
                 colorlinks = true,
                 linkcolor = BrickRed,
                 urlcolor  = blue,
                 citecolor = BrickRed,
                 anchorcolor = green,
                 hyperindex = true,
                 hyperfigures
                 ]{hyperref} 
                 
\newcommand{\comment}[1]{}

\newcommand{\sgn}{\mathrm{sign}}

\newcounter{figref}

\begin{document}

\title{\href{http://necsi.edu/research/social/foodprices.html}{The Food Crises: A quantitative model of food prices including speculators and ethanol conversion}} 
\date{\today}  
\author{Marco Lagi, Yavni Bar-Yam, Karla Z. Bertrand and \href{http://necsi.edu/faculty/bar-yam.html}{Yaneer Bar-Yam}}
\affiliation{\href{http://www.necsi.edu}{New England Complex Systems Institute} \\ 
238 Main St. Suite 319 Cambridge MA 02142, USA \vspace{2ex} \\ \footnotesize \emph{reviewed by:\\ C. Peter Timmer - Cabot Professor of Development Studies} emeritus. \emph{Harvard University} \\\emph{Jeffrey C. Fuhrer - Executive Vice President and Senior Policy Advisor. Federal Reserve Bank of Boston}\\\emph{Richard N. Cooper - Maurits C. Boas Professor of International Economics. Harvard University}\\\emph{Thomas C. Schelling - Distinguished Professor of Economics} emeritus. \emph{University of Maryland}\vspace{2ex}}

\begin{abstract}
Recent increases in basic food prices are severely impacting vulnerable populations worldwide. Proposed causes such as shortages of grain due to adverse weather, increasing meat consumption in China and India, conversion of corn to ethanol in the US, and investor speculation on commodity markets lead to widely differing implications for policy. A lack of clarity about which factors are responsible reinforces policy inaction. Here, for the first time, we construct a dynamic model that quantitatively agrees with food prices. The results show that the dominant causes of price increases are investor speculation and ethanol conversion. Models that just treat supply and demand are not consistent with the actual price dynamics. The two sharp peaks in 2007/2008 and 2010/2011 are specifically due to investor speculation, while an underlying upward trend is due to increasing demand from ethanol conversion. The model includes investor trend following as well as shifting between commodities, equities and bonds to take advantage of increased expected returns. Claims that speculators cannot influence grain prices are shown to be invalid by direct analysis of price setting practices of granaries. Both causes of price increase, speculative investment and ethanol conversion, are promoted by recent regulatory changes---deregulation of the commodity markets, and policies promoting the conversion of corn to ethanol. Rapid action is needed to reduce the impacts of the price increases on global hunger. 
\end{abstract}

\maketitle

\section{Food prices: Overview}
\label {sec:overview}

In 2007 and early 2008 the prices of grain, including wheat, corn and rice, rose by over 100\%, then fell back to prior levels by late 2008. A similar rapid increase occurred again in the fall of 2010. These dramatic price changes \cite{source_fao} have resulted in severe impacts on vulnerable populations worldwide and prompted analyses of their causes \cite{Timmer2008,Lustig2008,Baffes2010,Piesse2009,Fuglie2008,Gilbert2010,Abbott2009,AlRiffai2010,DeGorter2010,Benson2008,Alexandratos2008,Meyers2008,Beddington2010,Cooke2009,Clapp2010,Timmer2009,Hochman2008,Dewbre2008,Headey2010,Sekercioglu2008,Abbott2009_b,Sarris2009_1,Harrison2009,Searchinger2009_1,Lo2009,Tyner2010_1,Chen2011,Khanna2009,Saghaian2010,Wiggins2010,Armah2009,Baek2009,Chantret2009,Bureau2010,Sarris2009_2,Doering2009,Jansen2010,Martin2010,Pender2009,Orden2010,Timilsina2010_1,Kwon2009,Bouet,Rajagopal2009,Amponsah2009,Mondi2010,Tyner2010_2,Ahrens2010,Searchinger2009_2,Timilsina2010_2,Wright2010,Rutten2011,Rajcaniova2010,Meijerink2010,Rajagopal2010,Headey2009,Simelton2010,Abbott2010,Elliott2009,Rismiller2009,Singh2008,Perry2009,Onour2010,Clements2006}. Among the causes discussed are (a) weather, particularly droughts in Australia, (b) increasing demand for meat in the developing world, especially in China and India, (c) biofuels, especially corn ethanol in the US and biodiesel in Europe, (d) speculation by investors seeking financial gain on the commodities markets, (e) currency exchange rates, and (f) linkage between oil and food prices. Many conceptual characterizations and qualitative discussions of the causes suggest that multiple factors are important. However, quantitative analysis is necessary to determine which factors are actually important and which are not. While various efforts have been made, no analysis thus far has provided a direct description of the price dynamics. Here we provide a quantitative model of price dynamics demonstrating that only two factors are central: speculators and corn ethanol. 
We introduce and analyze a model of financial speculator price dynamics describing speculative bubbles and crashes. 
We further show that the increase in corn to ethanol conversion can account for the underlying price trends when we exclude speculative bubbles. A model combining both the shock due to increasing ethanol conversion and speculators quantitatively matches food price dynamics. Our results imply that changes in regulations of commodity markets that eliminated restrictions on investments \cite{CFMA,Kaufman2010,Marco-article,FedReg,Stewart2008}, and government support for ethanol production \cite{EnergyPolicyAct2005,Sandalow2007,Krigman2007,Hahn2008}, have played a direct role in global food price increases. 

The analysis of food price changes immediately encounters one of the central controversies of economics: whether prices are controlled by actual supply and demand, or are affected by speculators who can cause ``artificial" bubbles and panics. 
Commodity futures markets were developed to reduce uncertainty by enabling pre-buying or selling at known contract prices. In recent years ``index funds" that enable investors (speculators) to place bets on the increase of commodity prices across a range of commodities were made possible by market deregulation \cite{CFMA}. 
The question arises whether such investors, who do not receive delivery of the commodity, can affect market prices. One thread in the literature denies the possibility of speculator effects in commodities \cite{Lerner2000, Krugman_2011}. Others affirm a role for speculators in prices  \cite{Worthy2011,WEED_list,Timmer2008,Lustig2008,Baffes2010,Cooke2009,Kaufman2010,Piesse2009,Benson2008,Beddington2010,Alexandratos2008,Meyers2008,Timmer2009,Clapp2010,Rajagopal2009,Amponsah2009,Mondi2010}, but there has been no quantitative description of their effect. The rapid drop in prices in 2008, consistent with bubble/crash dynamics, increased the conviction that speculation is playing an important role. Still, previous analyses have been limited by an inability to 
directly model the role of speculators. This limitation has also been present  in historical studies of commodity prices. For example, analysis of sharp commodity price increases in the 1970s \cite{Cooper1975} found that they could not be due to actual supply and demand. The discrepancy between actual prices and the expected price changes due to consumption and production was attributed to speculation, but no quantitative model was provided for its effects. More recently, statistical (Granger) causality tests were used to determine whether any part of the price increases in 2008 could be attributed to speculative activity \cite{Robles2009,Cooke2009,Hernandez2010}. 
The results found statistical support for a causal effect, but the magnitude of the effect cannot be estimated using this technique.
A model of investors with ``bounded rational exuberance" in their investment strategies has shown increased price volatility, but has not been compared with actual price data \cite{Munier2010}.

Here we introduce a model relating speculation to prices and analyze its price dynamics. The model describes trend-following behavior and can directly manifest bubble and crash dynamics. In our model, when prices increase, trend following leads speculators to buy, contributing to further price increases. If prices decrease, the speculators sell, contributing to further price declines. Speculator trading is added to a dynamic model of supply and demand equilibrium. If knowledgeable investors believe supply and demand do not match (as inferred from available information), there is a countering (Walrasian) force toward equilibrium prices.  When prices are above equilibrium these investors sell, and when below these investors buy. 
The interplay of trend following and equilibrium restoring transactions leads to a variety of behaviors depending on their relative and absolute strengths. For a sufficiently large speculator volume, trend following causes prices to depart significantly from equilibrium. Even so, as prices further depart from equilibrium the supply and demand restoring forces strengthen and eventually reverse the trend, which is then accelerated by the trend following back toward and even beyond the equilibrium price. The resulting oscillatory behavior, consisting of departures from equilibrium values and their restoration, matches the phenomenon of bubble and crash dynamics. The model clarifies that there are regimes in which speculators have distinct effects on the market behavior, including both stabilizing and destabilizing the supply and demand equilibrium.

\begin{figure}[tb]
\refstepcounter{figref}\label{fig:food_fit}
\href{http://necsi.edu/research/social/img/fig1.pdf}{\includegraphics[width=0.9\linewidth]{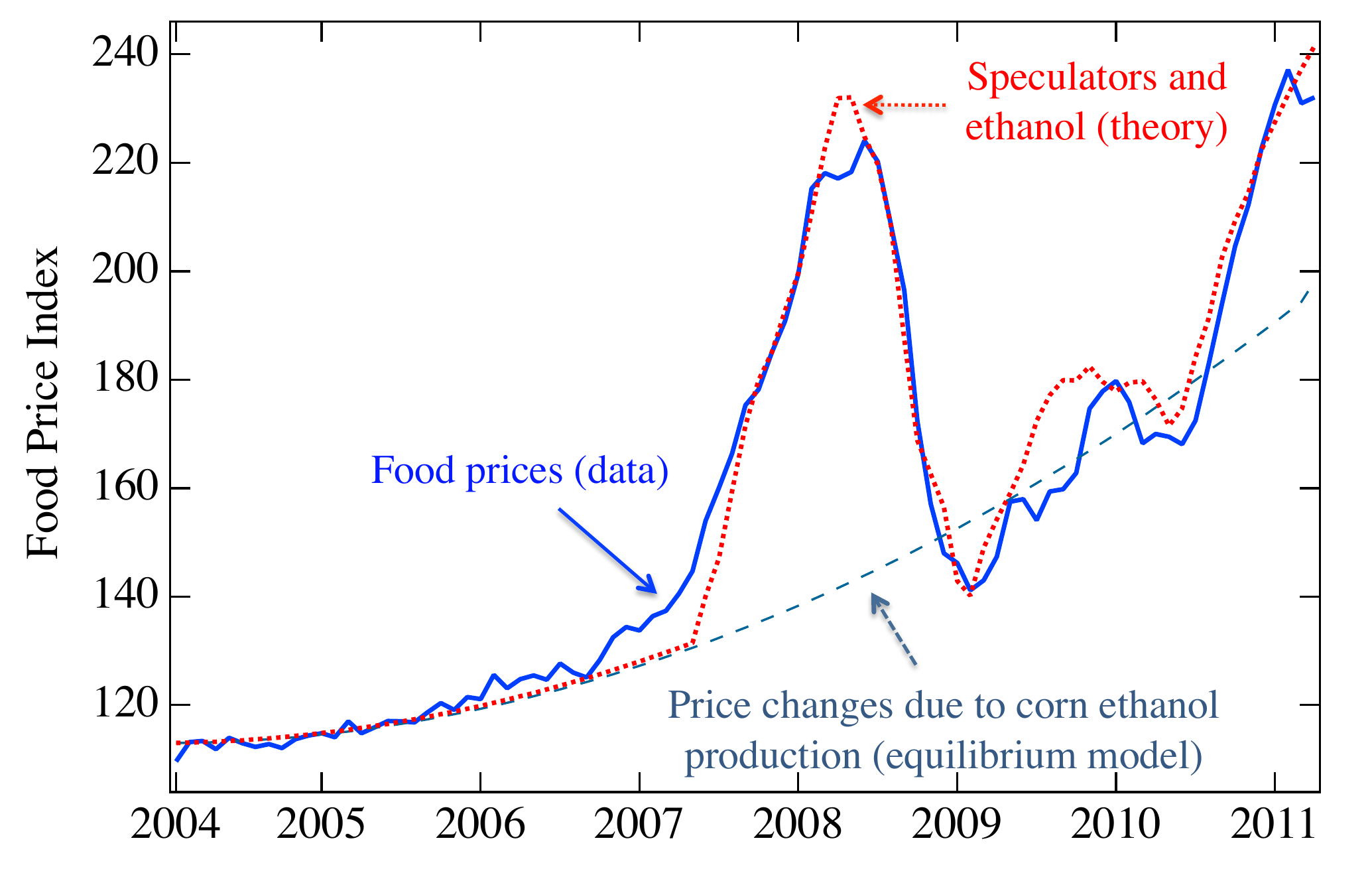}}
\caption{\textbf{Food prices and model simulations} - The FAO Food Price Index (blue solid line) \cite{source_fao}, the ethanol supply and demand model (blue dashed line), where dominant supply shocks are due to the conversion of corn to ethanol so that price changes are proportional to ethanol production (see Appendix \hyperref[app:c]{C}) and the results of the speculator and ethanol model (red dotted line), that adds speculator trend following and switching among investment markets, including commodities, equities and bonds (see Appendices \hyperref[app:d]{D} and \hyperref[app:e]{E}).}
\end{figure}

We further systematically consider other proposed factors affecting food prices. We provide quantitative evidence excluding all of them from playing a major role in recent price changes except corn to ethanol conversion. We show that, aside from the high price peaks, the underlying trends of increasing food prices match the increases in the rate of ethanol conversion. We construct a dominant supply shock model of the impact of ethanol conversion on prices, accurately matching underlying price trends and demonstrating that the supply and demand equilibrium prices would be relatively constant without the increase in corn to ethanol conversion. We then combine the effects of speculators and corn to ethanol conversion into a single model with remarkably good quantitative agreement with the food price dynamics. Final results are shown in Fig. \ref{fig:food_fit}. The unified model captures the way speculators shift between equities and commodities for maximum projected gains. 

In order to complete the picture, and respond to claims that commodity market investors/traders have no mechanism for influencing actual commodity (spot) market prices \cite{Krugman_2011}, we interviewed participants in the spot market who state unequivocally that they base current prices on the futures market \cite{Ronnie2011,Tanger2011}. The use of futures prices as a reference enables speculative bubbles on the futures market to influence actual food prices. 

\begin{figure}[t]
\refstepcounter{figref}\label{fig:stocks}
\centerline{
\href{http://necsi.edu/research/social/img/fig2.pdf}{\includegraphics[width=150mm]{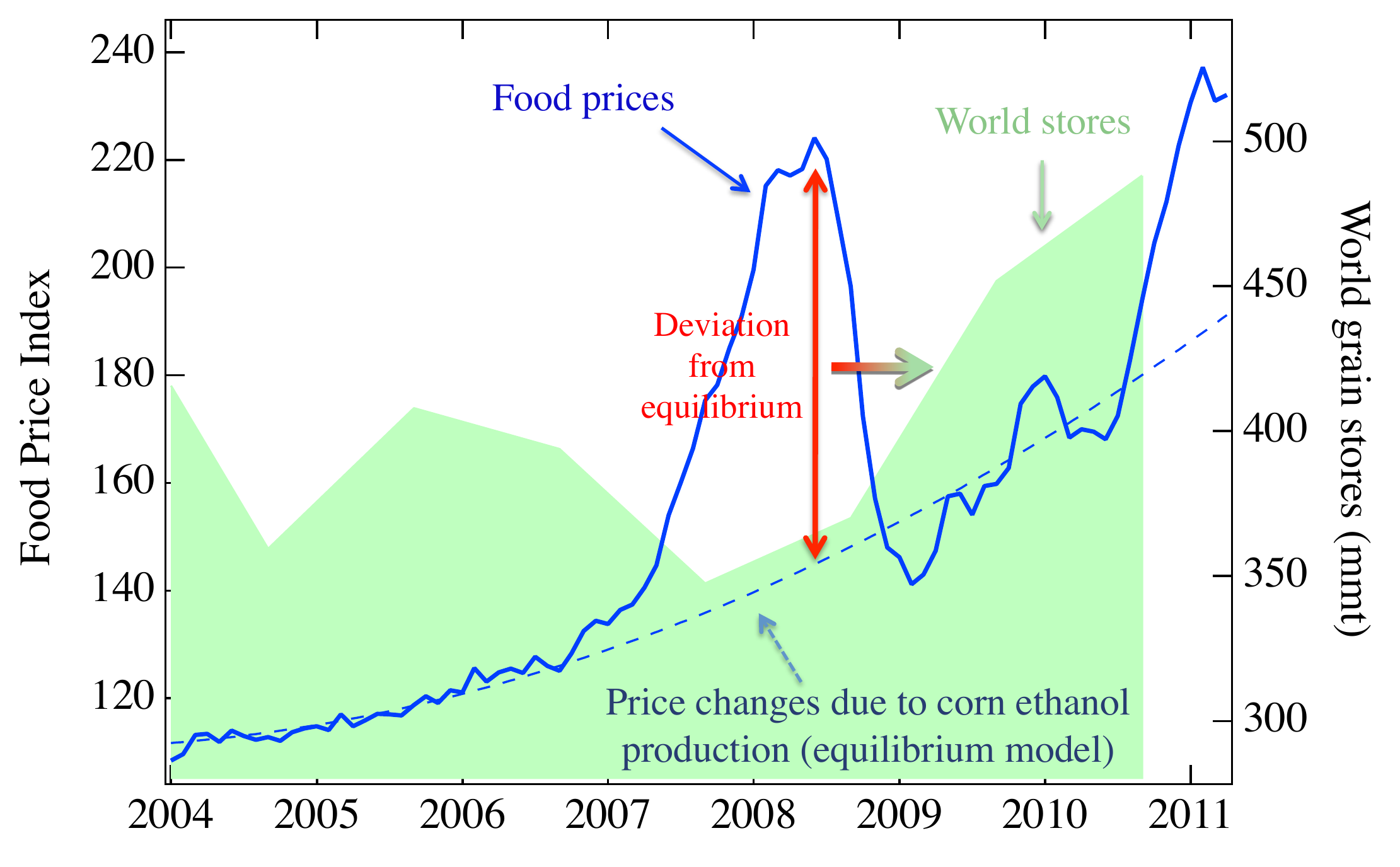}}
}
\caption{\textbf{Impact of food prices on grain inventories} - A deviation of actual prices (solid blue curve) from equilibrium (dashed blue curve) indicated by the red arrow leads to an increase in grain inventories (green shaded area) delayed by approximately a year (red to green arrow). This prediction of the theory is consistent with observed data for 2008/2009.  Increasing inventories are counter to supply and demand explanations of the reasons for increasing food prices in 2010. Restoring equilibrium would enable vulnerable populations to afford the accumulating grain inventories.}
\end{figure}

We can then consider how the deviations of food prices from equilibrium impact grain inventories, and how these in turn influence grain prices. Prices above equilibrium reduce demand and increase supply leading to accumulation of grain inventories. 
However, while prices affect decisions immediately, the very nature of futures contracts is that delivery occurs after contract maturation. Futures contracts may be bought with maturity horizons at intervals of three, six, nine and twelve months, or more. The expected time delay is the characteristic time over which producers and consumers choose to contract for delivery, reflecting their hedging and planning activities. Interviews of market participants suggest it can be reasonably estimated to be six months to a year due to both agricultural cycles and financial planning \cite{Wright2011, Hostetter2011}.  When prices are above their equilibrium values, and after this time delay, inventories would be predicted to increase due to high prices that reduce demand and increase supply. Thus, our model predicts that price deviations from equilibrium will be accompanied after a time delay by changes in grain inventories. Figure \ref{fig:stocks} shows that this prediction is consistent with empirical data \cite{source_grains}. World grain inventories increased most rapidly between Sept. 2008 and 2009, one year after the first speculative bubble. (Claims of decreasing inventories refer to the period before 2008 \cite{Sanders2008}.) 
Inventories continued to increase, but less rapidly, one year after the near equilibrium prices of 2009. According to the model, this period involved a rapid increase in corn use for ethanol production and shifting of food consumption to other grains, which was a major shock to the agriculture and food system. The increasing inventories are not consistent with supply and demand reasons for the price increases in 2010, but are consistent with our model in which the rising prices in 2010 are due to speculation. 

As inventories increase, inventory information will become available after an additional time delay. This information could influence investors, leading to the kind of Walrasian selling and buying that would reverse trends and restore equilibrium prices, i.e. cause a crash. The market reaction for pricing might be delayed further by the time participants take to react to these signals. Still, this provides an estimate of the duration of speculative bubbles. Indeed, the time until the peak of the bubbles of approximately 12 months in both 2007-8 and 2010-1 provides a better estimate of time frames than the coarser inventory data does and is consistent with the financial planning timeframes of producers and consumers. 
This suggests that investors may only be informed after actual supply and demand discrepancies are manifest in changing inventories. 
The existence of a second speculator bubble in 2010 raises the question of why speculators did not learn from the first crash to avoid such investing. Speculators, however, profited from the increase as well as lost from the decline and they may have an 
expectation that they can successfully time market directional changes, leaving others with losses (the ``greater fool theory"). 

The recent increasing inventories also raise humanitarian questions about the current global food crisis and efforts to address hunger in vulnerable populations in the face of increasing world prices \cite{hunger1,hunger2,hunger3,FAOreport}. 
The amount of the increase in inventories---140 million metric tons (mmt) from Sept 2007 to Sept 2010---is the amount consumed by 440 million individuals in one year. According to our model, the reason much of this grain was not purchased and eaten is the increase in food prices above equilibrium values due to speculation. This unconsumed surplus along with the 580 mmt of grain that was used for ethanol conversion since 2004 totals 720 mmt of grain, which 
could otherwise have been eaten by many hungry individuals. These outcomes are not only ethically disturbing, they are also failures of optimal allocation according to economic principles. The deregulation of commodity markets resulted in non-equilibrium prices that caused a supply and demand disruption/disequilibrium driving lower consumption and higher production---inventories
accumulated while people who could have afforded the equilibrium prices went hungry. Regulation of markets and government subsidies to promote corn to ethanol conversion have distorted the existing economic allocation by diverting food to energy use. This raised equilibrium prices, increased energy supply by a small fraction (US corn ethanol accounted for less than 1\% of US energy consumption in 2009 \cite{source_energy}) and reduced grain for food by a much larger one (US corn used for ethanol production is 4.3\% of the total world grain production, even after allowing for the feed byproduct \cite{source_grains,Hoffman2010}). The failures of both deregulation and regulation ably demonstrate that the central issue for policy is not whether to regulate, but how to choose the right regulations. 

Our results have direct implications for understanding the complex dependencies of global economics and the societal effects of food prices. The flows of capital in global markets can be traced from the financial crisis through our speculator model. Due to the collapse of the mortgage market and the stock market crash, investors moved money to the commodities market. This resulted in boom-bust cycles, including in food and other commodities. In a separate paper we describe the connection between food prices and the recent social unrest, violence and government changes in North Africa and the Middle East \cite{food_crises}. Our analysis extends the dominos of global interdependence from housing, to the stock market, to the commodities market, to social unrest. Policy discussions should recognize the extent of such links. Here we focus on the food prices and their causes.  

We divide our discussion into three parts: first the role of changes in supply and demand (Section \ref{sec:sd}), second the impact of speculator investment (Section \ref{sec:speculation}), and third the role of exchange rates and energy costs (Section \ref{sec:oil}). After the conclusions (Section \ref{sec:conclusion}), we provide a summary of prior studies in Appendix \hyperref[app:a]{A} and details about our quantitative models in Appendices \hyperref[app:b]{B}--\hyperref[app:e]{E}.

\section{Changes in supply and demand}
\label {sec:sd}

In order to account for the recent observed food price differences, changes in supply and demand would have to be much larger than normal variation and rapid enough to have impact over the period of a year. 
Candidates for the causal factors include weather affecting the supply, increasing consumption of meat and other livestock products in developing nations causing changes in demand, and the use of corn for ethanol production. 

\begin{figure}[htpb]
\refstepcounter{figref}\label{fig:explanations}
\href{http://necsi.edu/research/social/img/fig3.pdf}{\includegraphics[width=0.92\linewidth]{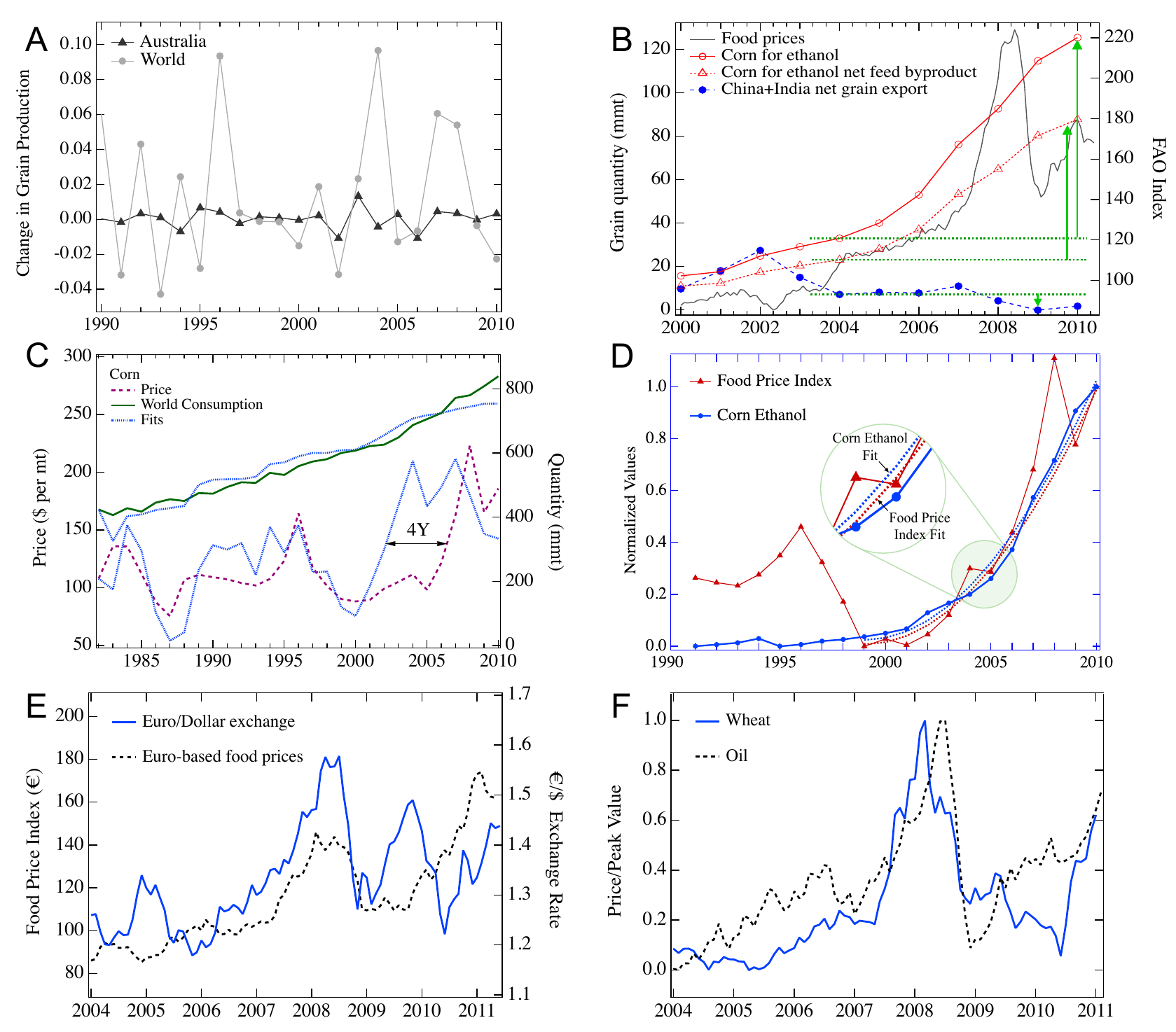}}
\caption{\textbf{Analysis of possible causes of food price increases} - \emph{A}: Weather, specifically droughts in Australia. Comparison of change in world (grey) and Australian (black) grain production relative to total world production by weight
\cite{source_grains}. The correlation is small. \emph{B}: Emerging markets, specifically meat consumption in China. Comparison of China and India net grain export (dashed blue) to the US corn ethanol conversion demand (solid red) and net demand after feed byproduct (dotted red) \cite{Hoffman2010}, and FAO food price index (solid black). Arrows show the maximum difference from their respective values in 2004. The impact of changes in China and India is much smaller. \emph{C}: Supply and demand. Corn price (dashed purple) and global consumption (solid green) along with best fits of supply and demand model (blue) (see Appendix \hyperref[app:b]{B}) \cite{source_grains}. Price is not well described after 2000. \emph{D}: Ethanol production (Fig. \ref{fig:fit_corn}). US corn used for ethanol production (blue circles) and FAO Food Price Index (red triangles). Values are normalized to range from 0 to 1 (minimum to maximum) during the period 1990-2010. Dotted lines are best fits for quadratic growth, with coefficients of $0.0083 \pm 0.0003$ and $0.0081 \pm 0.0003$ respectively. The 2007/8 bubble was not included in the fit or normalization of prices \cite{source_grains}. \emph{E}: Currency conversion: euro-based FAO Food Price Index (dashed black), euro/dollar exchange (solid blue) \cite{source_exchange}. Both have peaks at the same times as the food prices in dollars. However, food price increases in dollars should result from decreasing exchanges rates. \emph{F}: Oil prices. Wheat price (solid blue) and Brent crude oil price (dashed black). The peak in oil prices follows the peak in wheat prices and so does not cause it \cite{source_oil}.}
\end{figure}

The most common explanation provided by market news interpreters for the 2008 food price increases was the drought in Australia \cite{Bradsher2008,Bryant2008,Collier2007}. However, the production of grains in Australia does not correlate with global production (Fig. \ref{fig:explanations} A). The Pearson correlation coefficient of the two time series over the last 20 years is only $\rho=0.17$. Other countries have increases and decreases based upon variable conditions and therefore the changes in global production are not well described by Australia's production. The fraction of global grain production from Australia (circa 1.8\% by weight in 2010 \cite{source_grains}) is therefore not sufficient to be a significant causal factor at the magnitude of influence of recent price changes, even if it might be at smaller scales and shorter time frames. 
In particular, the low production in Australia in 2006 did not coincide with a global production decrease, and in 2007 both Australia and the world had increases in production (Fig. \ref{fig:explanations} A). Droughts in Australia, and global weather conditions more generally, are therefore unable to explain the recent food price changes. 

A widely cited potential longer term cause of increasing prices is a change of diet from grains to meat and other livestock products,
as a result of economic development \cite{Brown1995,Schelling2011}. Development of China, India, and other countries, comprising more than one-third of the world population, has created higher food demands as the diet of these countries changes. Changes in diet might have a large impact on the consumption of feed grains, as the ratio of animal feed to meat energy content has been estimated to be as high as 4:1, 17:1 and 50:1 for chicken, pork and beef respectively \cite{Cornell1997}. However, the increasing demand for grain in China and India has been met by internal production and these countries have not, in recent years, been major participants in the global grain markets \cite{source_grains}. Indeed, demand growth in these countries slowed in the years leading up to the food price spike in 2008 \cite{Alexandratos2008,Baffes2010}, and the countries combined remained net exporters \cite{Alexandratos2008,Abbott2009_b}. As shown in Fig. \ref{fig:explanations} B, their combined net international export of grains has decreased by 5 mmt, from 7 mmt in 2004 to 2 mmt in 2010 \cite{source_grains}. In contrast, the increase in the amount of corn used for ethanol production is 20 times larger, 95 mmt (if we subtract a feed byproduct of ethanol production \cite{Hoffman2010} it is 13 times larger, 67 mmt). 
The increase in demand due to corn feed in China, for all purposes but
primarily for hogs (the dominant source of meat), from 2004 to 2010 is 22 mmt, less than one-quarter of the ethanol demand (one third after feed byproduct). Even this amount was mostly met by internal production increases. Import and export policies isolate the Chinese domestic grain market and domestic prices
of feed grains do not track global prices, so only the reduction of net export affects the global market.  The impact on global food prices of changes in feed grain demand due to economic development is therefore negligible with respect to US demand for corn for ethanol. 

The many possible reasons for changes in supply and demand can be considered together if they result in a surplus or deficit, and this surplus or deficit is the primary reason for changes in grain inventories. Grain inventories can then be used as an indicator of supply and demand shocks to construct a quantitative model of prices \cite{Abbott2008}.
However, estimates of inventories provided by the US Department of Agriculture are not consistent with global food prices when considered within such a quantitative supply and demand model. In Appendix \hyperref[app:b]{B} fluctuations in wheat, corn and rice inventories and prices are treated in this way. The example of corn is shown in Fig. \ref{fig:explanations} C. The prices change due to either a supply shock or a demand shock given by the net surplus of a given commodity. Prices shift upwards if there is a deficit and downwards if there is a surplus. In principle, the model allows a fit of both the observed price of the commodity and its consumption (or production). Prior to 2000 the main features of price dynamics can be fit by the model, consistent with earlier studies on the role of supply and demand \cite{Sumner1989,Deaton1990}. However, since 2000, both the price and consumption values, including the recent large price increases, are not well described. There are reductions in the inventories around the year 2000, which give rise to significant price increases according to the model. However, the timing of these model-derived price increases precedes by three to four years the actual price increases. Also, the model implies an increase in consumption at that time that does not exist in the consumption data. Among the reasons for a reduction in reserves in 2000 is a policy change in China to decrease inventories \cite{Huang2008, Abbott2009}. Such a policy change would affect reserves but would not describe market supply and demand. Another reason for the inability for the supply and demand model to describe prices is the role of speculation in prices that leads to non-equilibrium prices and changes in grain inventories as discussed in the next section. The high peaks of recent price behavior have also suggested to some that the mechanism is a decline of supply and demand elasticities, i.e. high sensitivity of prices to small variations in supply and demand quantities \cite{Abbott2009}. However, for this explanation to be valid, supply and demand shocks must still correspond to price dynamics, and this connection is not supported in general by Granger causality analysis \cite{Timmer2008,Cooke2009}.

Finally, we consider conversion of corn to ethanol. Only a small fraction of the production of corn before 2000, corn ethanol consumed a remarkable 40\% of US corn crops in 2011 \cite{source_grains}, promoted by US government subsidies based upon the objective of energy independence \cite{EnergyPolicyAct2005,Sandalow2007,Krigman2007,Hahn2008}, and advocacy by industry groups \cite{Hahn2008,Wallen2010,Globe2009}. Corn serves a wide variety of purposes in the food supply system and therefore has impact across the food market \cite{GFO,Abbassian2006,Pollan2006}. Corn prices also affect the price of other crops due to substitutability at the consumer end and competition for land at the production end \cite{Timmer2008}. There have been multiple warnings of the impact of this conversion on global food prices and world hunger \cite{Tenenbaum2008,Mitchell2008,Rosegrant2008,Economist2007,Vidal2007,Chronicle2010,Weise2011,Walsh2011}, and defensive statements on the part of industry advocates \cite{Bennett2011,Block2011}. Among quantitative studies (Appendix \hyperref[app:a]{A}), ethanol conversion is most often considered to have been the largest factor in supply and demand models. Absent a model of speculators, ethanol conversion is sometimes considered the primary cause of price increases overall (Appendix \hyperref[app:a]{A}). However, ethanol conversion itself cannot describe the dynamics of prices because ethanol production has been increasing smoothly since 2004.  Therefore, it cannot explain the sharp decline of prices in 2008. 
We show that ethanol can account for the smoothy rising prices once the high peaks are accounted for by speculation. Fig. \ref{fig:explanations} D compares annual corn ethanol production and food prices. During the period 1999-2010, ignoring the 2007-2008 peak, the two time series can be well fitted by the same quadratic growth (no linear term is needed). The quadratic coefficients are $0.0083 \pm 0.0003$ for corn ethanol and $0.0081 \pm 0.0003$ for food prices, which are the same within fitting uncertainty. The quality of the fits is outstanding, with $R^2$ values of $0.986$ and $0.989$ respectively. The Pearson correlation coefficient of the food price and ethanol annual time series is $\rho=0.98$. The parallel increase of the two time series since 2004 suggests that corn ethanol is likely to be responsible for the underlying increase in the cost of food during this period. As shown in Appendix \hyperref[app:c]{C} the relationship between food prices and corn to ethanol conversion can be obtained by modeling the impact of corn ethanol production as a dominant shock to the agricultural system. According to this model, other supply and demand factors would leave the prices mostly unchanged. Prior to 1999 corn ethanol production and prices are not correlated because of the small amount of ethanol production. Price variation during that period must be due to other causes.  

\section{Speculation}
\label {sec:speculation}

The role of speculation in commodity prices has been considered for many years by highly regarded economists \cite{Cooper1975,WEED_list}. There is a long history of speculative activity on commodity markets and regulations were developed to limit its effects \cite{Speculation1947,Markham1986,Markham1991}. Recently, however, claims have been made that there is no possibility of speculator influence on commodity prices because investors in the futures market do not receive commodities \cite{Krugman_2011, Lerner2000}. We have  
investigated this claim by asking individuals who set prices at granaries (the spot market) and who monitor the  prices at the US Department of Agriculture how they determine the prices at which to buy or sell \cite{Ronnie2011,Tanger2011}. They state that spot market prices are set according to the Chicago Board of Trade futures exchange, assuming that it reflects otherwise hidden global information, with standard or special increments to incorporate transportation costs, profits, and when circumstances warrant, slight changes for over- or under-supply at a particular time in a granary. Thus the futures market serves as the starting point for spot market prices. The conceptual temporal paradox of assigning current prices based upon futures is not considered a problem, and this makes sense because grains can be stored for extended periods. 
 
\begin{figure}[t]
\refstepcounter{figref}\label{fig:bubbleprotests}
\centerline{
\href{http://necsi.edu/research/social/img/fig4.pdf}{\includegraphics[width=150mm]{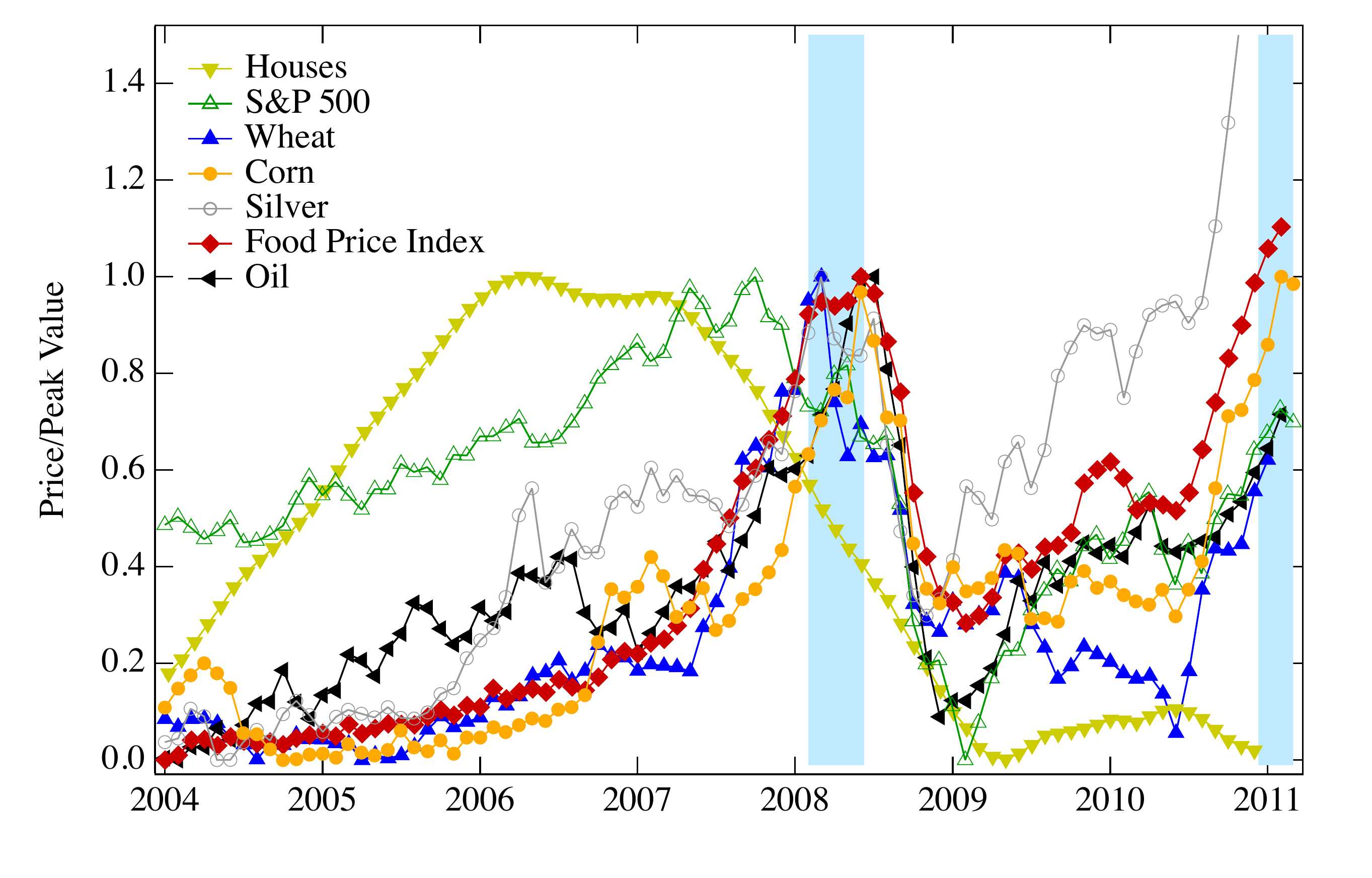}}
}
\caption{\textbf{Time dependence of different investment markets} - Markets that experienced rapid declines, ``the bursting of a bubble," between 2004 and 2011. Houses (yellow) \cite{source_houses}, stocks (green) \cite{source_stocks}, agricultural products (wheat in blue, corn in orange) \cite{source_grains}, silver (grey) \cite{source_oil}, food (red) \cite{source_fao} and oil (black) \cite{source_oil}. Vertical bands correspond to periods of food riots and the major social protests called the ``Arab Spring" \cite{food_crises}. Values are normalized from 0 to 1, minimum and maximum values respectively, during the period up to 2010.}
\end{figure}

If commodities futures investors determine their trading based upon supply and demand news, the use of the futures market to determine spot market prices, discounting storage costs, would be a self-consistent way of setting equilibrium prices \cite{Hotelling1931,Stein1961,Pinkdyck2001}. But if investors are ineffective in considering news or are not motivated by supply and demand considerations, deviations from equilibrium and speculative bubbles are possible. When prices depart from equilibrium, accumulation or depletion of inventories may result in an equilibrium restoring force. This impact is, however, delayed by market mechanisms. Since producers and consumers generally hedge their sales and purchases through the futures market, transactions at a particular date may immediately impact food prices and decisions to sell and buy, but impact delivery of grains at a later time when contracts mature. The primary financial consequences of a deviation of prices from equilibrium do not lead to equilibrium-restoring forces. Producers, consumers and speculators each have gains and losses relative to the equilibrium price, depending on the timing of their transactions, but the equilibrium price is not identified by the market. Profits (losses) are made by speculators who own futures contracts as long as futures prices are increasing (decreasing), and by producers as long as the prices are above (below) equilibrium. When prices are above equilibrium consumers incur higher costs which may reduce demand. Producers may increase production due to higher expected sales prices. The result of this reduction and increase is an expected increase in inventories after a time delay: an agricultural or financial planning cycle, which may be estimated to be six months to a year \cite{Wright2011, Hostetter2011}.
Finally, the feedback between increased inventories and price corrections requires investors to change their purchases. First the information about increased inventories must become available. Even with information about increasing inventories, the existence of high futures prices can be interpreted as a signal of increased future demand, further delaying market equilibration.
Speculatively driven bubbles can thus be expected to have a natural duration of a year or longer (see Fig. \ref{fig:bubbleprotests}). 
(We note that it is possible to relate trend following speculators to the ``supply of storage'' concept in which current inventories increase due to higher expected future prices \cite{sos_1,sos_2}. However, in doing so we encounter paradoxes of recursive logic, see Appendix \hyperref[app:d]{D}.)

We review the empirical evidence for the role of speculation in food prices, which includes the timing of the food price spikes relative to the global financial crisis, the synchrony of food price spikes with other commodities that do not share supply and demand factors, the existence of large upwards and downwards movement of prices consistent with the expectations of a bubble and bust cycle, statistical causality analysis of food prices increasing with commodity speculator activity, and an inability to account for the dynamics of prices with supply and demand equations despite many economic analyses. We add to these an explicit model of speculator dynamics which quantitatively fits the price dynamics. 

The mechanisms of speculator-driven food price increases can be understood from an analysis of the global consequences of the financial crisis. This analysis connects the bursting of the US real estate market bubble and the financial crisis of 2007-2008 to the global food price increases \cite{Caballero2008,crisis_impact_report}. Figure \ref{fig:bubbleprotests} shows the behavior of the mortgage market (housing prices), stock market (S\&P 500), and several commodities: wheat, corn, silver, oil, and the FAO food price index. The increase in food prices coincided with the financial crisis and followed the decline of the housing and stock markets. An economic crisis would be expected to result in a decrease in commodity prices due to a drop in demand from lower overall economic activity. The observed counterintuitive increase in commodity prices can be understood from the behavior expected of investors in the aftermath of the collapse of the mortgage and stock markets: shifting assets to alternative investments, particularly the commodity futures market \cite{speculators_spiegel, Macwhirter2008, Wahl2008}. This creates a context for intermittent bubbles, where the prices increase due to the artificial demand of investment, and then crash due to their inconsistency with actual supply and demand, only to be followed by another increase at the next upward fluctuation. The absence of learning behavior can be explained either by the ``greater fool theory,'' whereby professionals assume they can move their assets before the crash and leave losses to less skilled investors, or by the hypothesis that traders are active for just one price cycle, and that the next cycle will see new traders in the market. Even without a quantitative analysis, it is common to attribute rapid drops in prices to bubble and crash dynamics because the rapid upwards and downward movements are difficult to reconcile with normal fundamental supply and demand factors \cite{Doggett2011,Timmer2008,Bernanke2010}.

In addition to the timing of the peak in food prices after the stock market crash, the coincidence of peaks in unrelated commodities including food, precious and base metals, and oil indicates that speculation played a major role in the overall increase \cite{Tang2011}. An explanation of the food price peaks in 2008 and 2011 based upon supply and demand must not only include an explanation of the rise in prices of multiple grains, including wheat, corn and rice, but must separately account for the rise in silver, oil and other prices. In contrast, speculator-driven commodity bubbles would coincide after the financial crisis because of the synchronous movement of capital from the housing and stock markets to the commodity markets. Moreover, the current dominant form of speculator investment in commodity markets is in index funds \cite{Worthy2011}, which do not differentiate the behavior of different commodities, as they are aggregate bets on the overall commodity market price behavior. Such investor activity acts in the same direction across all commodities, without regard to their distinct supply and demand conditions. The relative extent to which each type of commodity is affected depends on the weighting factors of their representation in index fund investing activity compared to the inherent supply and demand related market activity. 

Recently, the growth of commodity investment activity has been studied in relation to commodity prices \cite{Robles2009,Cooke2009,WEED_list,Timmer2008}. Since index fund investments are almost exclusively bets on price increases (i.e.\ ``long" rather than ``short" investments), the investment activity is an indication of pressure for price increases. Increases in measures of investment have been found to precede the increases in prices in a time series (Granger) causality analysis \cite{Robles2009,Cooke2009}. (An OECD study claiming that speculation played no role \cite{Irwin2010, Buttonwood2010}, has been discounted due to invalid statistical methods \cite{Frenk2010}.) Granger causality tests also show the influence of futures prices on spot market prices \cite{Hernandez2010}. The causality analysis results provide statistical evidence of a role of speculative activity in commodity prices. However, they do not provide quantitative estimates of the magnitude of the influence. 

For many analyses, the absence of a manifest change in supply and demand that can account for the large changes in prices is considered strong evidence of the role of speculators. As we described in the previous section, supply and demand analyses of grain prices do not account for the observed dynamics of price behavior. None of the causes considered, individually or in combination, have been found to be sufficient. Appendix \hyperref[app:a]{A} reviews multiple efforts which have not been able to fit the changes in food prices to fundamental causes. As with analyses of commodity price changes in relation to supply and demand in the 1970s, such an absence is evidence of the role of speculators \cite{Cooper1975}. 

In Appendix \hyperref[app:d]{D} we construct a quantitative model of speculator activity in the commodity futures market by directly considering the role of trend-following investor dynamics. Trend following results in an increase in investment when prices are rising, and a decrease when prices are declining. Our results describe bubble and crash dynamics when certain relationships hold between the amount of speculative investment activity and the elasticity of supply and demand. The resulting price oscillations can be modified by investors switching between markets to seek the largest investment gains. We use the model of speculators to describe their impact on the food price index. When we include trend following, market switching behaviors, and the supply and demand effects of changes in corn to ethanol conversion, the results, shown in Fig. \ref{fig:food_fit}, provide a remarkably good fit of the food price dynamics (Appendix \hyperref[app:e]{E}). We find the timescale of speculative bubbles to be $11.8$ months, consistent with annual financial planning cycles and the maturation of futures contracts for delivery \cite{Wright2011, Hostetter2011}.
While there have been no such direct models that match observed price dynamics, trend following has been analyzed theoretically as a mechanism that can undermine fundamental price equilibrium \cite{DeLong1990_a,DeLong1990_b}, and is a central component of actual investing: advisors to commodity investors provide trend-following software and market investment advice based upon ``technical analysis" of time series \cite{CRB}. Such market investment advice does not consider weather or other fundamental causes. Instead it evaluates trends of market prices and their prediction using time series pattern analysis. Trend following is also the core of the recently proposed formalization of ``bounded rational exuberance"  \cite{Munier2010}.

We note that our analysis of the effect of commodity investments on the food price index aggregates the impact of speculator investment across multiple grains. However, it is enlightening to consider the impact on the rice price dynamics in particular. The direct impact of speculators on rice is small because rice is not included in the primary commodity index funds, as it is not much traded on the US exchanges. Instead, the price of rice is indirectly affected by the prices of wheat and corn, especially in India where wheat and rice can be substituted for each other. A sharp price peak in rice occurred only in 2008 (there is no peak in 2010) and this peak can be directly attributed to the global reaction to India's decision, in the face of rising wheat prices, to stop rice exports \cite{Timmer2008,Timmer2009_tr,Meyers2008}. The observation that rice did not have the behavior of other grains is consistent with and reinforces our conclusions about the importance of speculators in the price of corn and wheat, and thus food overall. 

\section{Additional factors: exchange rates and energy costs}
\label {sec:oil}

Two additional factors have been proposed to have a causal role in food prices: currency exchange and energy prices. 

Dollar to euro conversion rates are, at times, correlated to commodity prices \cite{Timmer2008, Abbott2008}. During these periods an increase in commodity prices coincides with an increase in euro value relative to the dollar. It has been suggested that the reason that food prices increased in dollars is because commodities might be priced primarily in euros, which would cause prices to rise in dollars. This has been challenged on a mechanistic level due to the dominance of dollars as a common currency around the world and the importance of the Chicago futures market (CBOE) \cite{Tyner2010}. However, more directly, such a causal explanation is not sufficient, since the prices of commodities in euros have peaks at the same times as those in dollars, as shown in Fig. \ref{fig:explanations} E.  Since the US is a major grain exporter, a decline in the dollar would give rise to a decrease in global grain prices. 
(The effect is augmented by non-US grain exports that are tied to the dollar, and moderated by supply and demand corrections, but these effects leave the direction of price changes the same.)
The opposite is observed. Moreover, the exchange rate also experienced a third peak in 2009, between the two food price peaks in 2008 and 2011. There is no food price peak either in euros or dollars in 2009. This suggests that the correlation between food prices and exchange rates is not fundamental but instead may result from similar causal factors. 

Some researchers have suggested that increasing energy prices might have contributed to the food prices \cite{Piesse2009,Abbott2009_b,Tyner2010,Mitchell2008}. This perspective is motivated by three observations: the similarity of oil price peaks to the food price peaks; the direct role of energy costs in food production and transportation; and the possibility that higher energy prices might increase demand for ethanol. Careful scrutiny, however, suggests that energy costs cannot account for food price changes. First, the peak of oil prices occurred {\em after} the peak in wheat prices in 2008, as shown in Fig. \ref{fig:explanations} F. Second, US wheat farm operating costs, including direct energy costs and indirect energy costs in fertilizer, increased from \$1.78 per bushel in 2004 to  \$3.04 per bushel in 2008 \cite{source_farms}. The increase of \$1.26, while substantial, does not account for the \$4.42 change in farmer sales price.  More specifically, the cost of fertilizers was about 5\% the total value of wheat (the value of the global fertilizer market was \$46 billion in 2007 \cite{fertilizer_guide}, 15\% of which was used for wheat \cite{Heffer2009}; the value of the global wheat market was \$125 billion \cite{source_grains, source_oil}). Third, the demand from ethanol conversion (Fig. \ref{fig:explanations} D) has increased smoothly over this period and does not track the oil price (see Fig. \ref{fig:explanations} F and Fig. \ref{fig:bubbleprotests}). The connection between oil prices and food prices is therefore not the primary cause of the increase in food prices. Indeed, the increased costs of energy for producers can be seen to be an additional effect of speculators on commodity prices. As shown in Figure \ref{fig:bubbleprotests}, a large number of unrelated commodities, including silver and other metals, have a sharp peak in 2008. Given that some of the commodities displayed cannot be linked to each other by supply and demand consideration (i.e. they are not complements or substitutes, and do not have supply chain overlaps), the similarity in price behavior can be explained by the impact of speculators on all commodities. Metal and agricultural commodity prices behave similarly to the energy commodities with which they are indexed \cite{Frenk2010}. It might be supposed that the increased cost of energy should be considered responsible for a portion of the increase in food prices. However, since the increases in production cost are not as large as the increases in sales price, the increase in producer profits eliminate the necessity for cost pass-through. The impact of these cost increases would not be so much directly on prices, but rather would moderate the tendency of producers to increase production in view of the increased profits.

\section{Conclusions and implications}
\label {sec:conclusion}

A parsimonious explanation that accounts for food price change dynamics over the past seven years can be based upon only two factors: speculation and corn to ethanol conversion. We can attribute the sharp peaks in 2007/2008 and 2010/2011 to speculation, and the underlying upward trend to biofuels. The impact of changes in all other factors is small enough to be neglected in comparison to these effects. Our analysis reinforces the conclusions of some economic studies that suggest that these factors have the largest influence \cite{Timmer2008,Tsioumanis2009}. Our model provides a direct way to represent speculators, 
test if they can indeed be responsible for price effects, and determine the magnitude of those effects. Our background check of the pricing mechanisms of the spot food price market confirms that futures prices are the primary price-setting mechanism, and that the duration of commodity bubbles is consistent with the delay in supply and demand restoring forces. Despite the artificial nature of speculation-driven price increases, the commodities futures market is coupled to actual food prices, and therefore to the ability of vulnerable populations---especially in poor countries---to buy food \cite{imf_report,Wahl2008,GHI2008,GHI2009,GHI2010}. 

Addressing the global food price problem in the short and long term is likely to require intentional changes in personal and societal actions. Over the longer term many factors and actions can play a role. Our concern here is for the dramatic price increases in recent years and the changes in supply and demand and investment activity that drove these price increases. The immediate implications of our analysis are policy recommendations for changes in regulations of commodity markets and ethanol production.

The function of commodity futures markets is benefitted by the participation of traders who increase liquidity and stabilize prices \cite{Mill1848,ref:Friedman_1953}. Just as merchants improve the distribution of commodities in space, traders do so over time. And yet, the existence of traders has been found to cause market behaviors that are counter to market function, resulting in regulations including the Commodity Exchange Act of 1936 \cite{CEA}. Arguments in favor of deregulation have cited the benefits that traders provide and denied other consequences, eventually resulting in deregulation by the Commodity Futures Modernization Act of 2000 \cite{CFMA}. Our results demonstrate the nonlinear effects of increased trader participation \cite{Baumol1957}. Higher than optimal numbers of traders are susceptible to bandwagon effects due to trend following that increase volatility and cause speculative bubbles \cite{Leibenstein1950}, exactly counter to the beneficial stabilizing effects of small numbers of traders. Since intermediate levels of traders are optimal, regulations are needed and should be guided by an understanding of market dynamics. These regulations may limit the amount of trading, or more directly inhibit bandwagon effects by a variety of means. Until a more complete understanding is available, policymakers concerned with the global food supply should restore traditional regulations, including the Commodity Exchange Act. Similar issues arise in the behavior of other markets, including the recent repeal of transaction rules (the uptick rule) that inhibited bandwagon effects in the stock market \cite{Harmon2008}.

Today, the economics of food production is directly affected by nationally focused programs subsidizing agricultural production in the US and other developed countries to replace fossil fuels. These policies impact global supply and demand and reflect local and national priorities rather than global concerns. Our analysis suggests that there has been a direct relationship between the amount of ethanol produced and (equilibrium) food price increases. Moderating these increases can be achieved by intermediate levels of ethanol production. Under current conditions, there is a tradeoff between ethanol production and the price of food for vulnerable populations. Since the ethanol market has been promoted by government regulation and subsidy, deregulation may be part of the solution. Alternative solutions may be considered, but in the short term, a significant decrease in the conversion of corn to ethanol is warranted. 

These policy options run counter to large potential profits for speculators and agricultural interests, and the appealing cases that have been made for the deregulation of commodity markets and for the production of ethanol. In the former case, the misleading arguments in favor of deregulation are not supported by the evidence and our analysis. Similarly, the influence of economic interests associated with the agricultural industry is reinforced by since-debunked claims of the role of ethanol conversion in energy security and the environment \cite{Hahn2008}.  Thus, a very strong social and political effort is necessary to counter the deregulation of commodities and reverse the growth of ethanol production. A concern for the distress of vulnerable populations around the world requires actions either of policymakers or directly of the public and other social and economic institutions. 

\section{Acknowledgments}
\label {sec:ackn}

We thank Kawandeep Virdee for help with the literature review, Anzi Hu, Blake Stacey and Shlomiya Bar-Yam for editorial comments, Rick Tanger for help with data sources, Peter Timmer, Jeffrey Fuhrer, Richard Cooper and Tom Schelling for reviews, and Homi Kharas for helpful comments on the manuscript. This work was supported in part by the Army Research Office under grant \#W911NF-10-1-0523. 


\newpage
\phantomsection
\label{app:a}
\begin{center}
\Large Appendix A\\
\Large Literature Review
\end{center}

The literature on the mechanisms of food price volatility is extensive \cite{Schultz1945,Newbery1981,Timmer1989,Williams1991,Timmer1995,Timmer2000,WorldBank2005,Rashid2008,Timmer2010,Gilbert2010_a,Naylor2010,Dawe2010}. In this appendix we summarize a sample of the literature on the causes of the food price crisis of 2007-2008. An earlier summary can be found in Ref. \cite{Abbott2008}. For each paper, we note which of several potential factors the authors examine: the change of diet in developing countries, biofuel conversion, financial speculation in the commodity futures market, the price of crude oil, and variation in currency exchange rates. We also list other possible causes addressed in each paper, and specify the timeframe in question, in particular whether it addresses just the rising prices in 2007/08, includes the subsequent decline and if it also includes the increase in 2010/11. For each potential factor, we indicate whether the the paper suggests or determines it to be a cause (``yes" or ``no"). We also specify whether the analysis presented in the paper is quantitative (bold, with asterisk ``*"), qualitative (italic), or only a passing mention (normal). If the paper does not consider a particular factor, that column is left blank.

\begin{sidewaystable} [h] \scriptsize
\center
	\begin{tabular}{| c | p{2.5cm} | c | c | c | c | p{2.5cm} | p{6cm} | l |}	
	\hline
\emph{\textbf{Paper}}	 &	\emph{\textbf{Change of diet (meat)}}	 &	\emph{\textbf{Weather}}	 &	\emph{\textbf{Biofuels}}	 &	\emph{\textbf{Speculation}}	 &	\emph{\textbf{Oil}}	 &	\emph{\textbf{Currency \mbox{exchange}}}	 &	\emph{\textbf{Other causes}}	 & \emph{\textbf{Time range}}	\\ 
\hline
\cite{Timmer2008}	 & no	 &	yes (wheat)	 &	yes	 &	\textbf{*yes} (no rice)	 &	yes	 &	\textbf{*at times} (via speculators)	 &	trade policies, thin market, panic (rice),  depletion of inventories, price transmission between commodities	 &	2007-8 rise \& fall	\\
\cite{Lustig2008}	&	no	&	no	&	yes	&	yes	&	yes	&	yes	&	trade policies	&	2007-8 rise \& fall	\\
\cite{Baffes2010}	 & \textbf{*no} &		 &	yes	 &	yes	 &	\textbf{*yes} ($\sim 30$\%)	 &	\emph{yes}	 &	fiscal expansion, lax monetary policy	 &	2007-8 rise \& fall	\\
\cite{Piesse2009}	 & &	\textbf{*no}	 &	yes	 &	yes	 &	yes	 &	maybe	 &	trade policies, thin market (rice), R\&D decline	 &	2007-8 rise	\\
\cite{Fuglie2008}	 & \emph{yes} &	\emph{yes}	 &	\emph{yes}	 &		 &		 &		 &	production decline: \textbf{*no}	 &	2007-8 rise	\\
\cite{Gilbert2010}	 & no	 &		 &	\textbf{*yes}	 &	maybe	 &	no	 &	\textbf{*no}	 &		 &	2007-8 rise \& fall	\\
\cite{Abbott2009}	 & \textbf{*no} &	yes (wheat)	 &	yes	 &	maybe	 &	yes	 &	yes	 &	trade policies, depletion of inventories
	 &	2007-8 rise \& fall	\\
\cite{AlRiffai2010}	 & &		 &	yes	 &		 &		 &		 &		 &	2007-8 rise	\\
\cite{DeGorter2010}	 & &		 &	\textbf{*yes}	 &		 &		 &		 &		 &		\\
\cite{Benson2008}	 & \emph{yes} &	\emph{yes}	 &	\emph{yes}	 &	\emph{yes}	 &	\emph{yes}	 &	\emph{yes}	 &	climate change, productivity and R\&D decline, trade policies	 &	2007-8 rise\\
\cite{Alexandratos2008}	 & \textbf{*no} &	\emph{yes}	 &	\textbf{*yes}	 &	yes	 &	\emph{yes}	 &	maybe	 &	trade policies, population increase	 &	2007-8 rise \& fall	\\
\cite{Meyers2008}	 & no &	yes	 &	yes	 &	yes	 &	yes	 &	yes	 &	trade policies (rice), depletion of inventories &	2007-8 rise	\\
\cite{Beddington2010}	 & yes &	yes	 &	yes	 &	yes	 &	yes	 &	yes	 &	trade policies, population increase	 &	2007-8 rise	\\
\cite{Cooke2009}	 & maybe &	maybe	 &	maybe	 &	\textbf{*yes}	 &	\emph{yes}	 &	maybe	 &		 &	2007-8 rise	\\
\cite{Clapp2010}	 & &		 &		 &	yes	 &		 &		 &		 &	2007-8 rise	\\
\cite{Timmer2009}	 & &		 &		 &	\textbf{*yes}	 &		 &		 &		 &	2007-8 rise	\\
\cite{Hochman2008}	 & &		 &	\textbf{*yes}	 &		 &		 &		 &		 &		\\
\cite{Dewbre2008}	 & &	\emph{yes}	 &	\textbf{*yes}	 &		 &	yes	 &	yes and no	 &	income growth	 &		\\
\cite{Headey2010}	 & yes (long term) &	yes (long term)	 &	yes (long term)	 &	maybe	 &	yes	 &	yes (long term)	 &	trade policies	 &	2007-8 rise	\\
\cite{Sekercioglu2008}	 & &		 &	no	 &		 &	yes	 &		 &		 &		\\
\cite{Abbott2009_b}	 & no &	yes (wheat)	 &	yes	 &	maybe	 &	yes	 &	no	 &	depletion of inventories, trade policies	 &	2007-8 rise \& fall	\\
\cite{Sarris2009_1}	 & no &		 &	no	 &	no	 &	yes	 &	yes	 &	countries hoarding, trade policies	 &	2007-8 rise \& fall	\\
\cite{Harrison2009}	 & &		 &	yes	 &		 &		 &		 &		 &	2007-8 rise	\\
\cite{Searchinger2009_1}	 & &		 &	yes	 &		 &		 &		 &		 &		\\
\cite{Lo2009}	 & \emph{yes} &	\emph{yes}	 &	\emph{yes}	 &		 &	yes	 &		 &	productivity and R\&D decline	 &	2007-8 rise	\\
\cite{Tyner2010_1}	 & &		 &	\textbf{*yes}	 &		 &	\textbf{*yes} (via biofuels)	 &		 &		 &	2007-8 rise \& fall	\\
\cite{Chen2011}	 & &		 &	yes	 &		 &		 &		 &		 &	2007-8 rise	\\
\cite{Khanna2009}	 & &		 &	yes	 &		 &		 &		 &		 &	2007-8 rise	\\
\cite{Saghaian2010}	 & &		 &		 &		 &	maybe	 &		 &		 &	2007-8 rise	\\
\cite{Wiggins2010}	 & &	yes (short term)	 &	yes (short term)	 &		 &	yes (short term)	 &	yes (long term)	 &		 &	2007-8 rise \& fall	\\
\cite{Armah2009}	 & yes &		 &	yes	 &		 &	yes	 &	yes	 &		 &	2007-8 rise	\\
\cite{Baek2009}	 & &		 &		 &		 &	\textbf{*yes} (via biofuels)	 &	\textbf{*yes}	 &		 &		\\
	\hline
	\end{tabular}
\caption{Literature review, part $1$. See text for notation.}
\label{tab:biblio1}
\end{sidewaystable}

\begin{sidewaystable} [h] \scriptsize
\center
	\begin{tabular}{| c | p{2.5cm} | c | p{3cm} | p{3cm} | p{1cm} | p{2.5cm} | p{6cm} | p{3cm} |}	
	\hline
\emph{\textbf{Paper}}	 &	\emph{\textbf{Change of diet (meat)}}	 &	\emph{\textbf{Weather}}	 &	\emph{\textbf{Biofuels}}	 &	\emph{\textbf{Speculation}}	 &	\emph{\textbf{Oil}}	 &	\emph{\textbf{Currency \mbox{exchange}}}	 &	\emph{\textbf{Other causes}}	 & \emph{\textbf{Time range}}	\\ 
\hline
\cite{Chantret2009}	 & &		 &	yes	 &		 &	yes	 &		 &		 &	2007-8 rise \& fall	\\
\cite{Bureau2010}	 & &		 &	yes	 &		 &		 &		 &		 &		\\
\cite{Sarris2009_2}	&		&		&	yes	&	no	&	yes	&		&		&	2007-8 rise	\\
\cite{Doering2009}	&		&		&	maybe	&		&		&		&		&	2007-8 rise	\\
\cite{Jansen2010}	&		&		&		&		&		&		&	financial crisis	&	2007-8 fall	\\
\cite{Martin2010}	&		&		&	yes	&		&	yes	&	yes	&	recession	&	2007-8 rise \& fall	\\
\cite{Pender2009}	&		&	yes	&		&		&	yes	&	yes	&	land degradation	&	2007-8 rise	\\
\cite{Orden2010}	&		&		&		&		&		&		&	monetary policy	&	2007-8 rise \& fall	\\
\cite{Timilsina2010_1}	&		&		&	yes	&		&		&		&		&	2007-8 rise	\\
\cite{Kwon2009}	&		&		&		&		&	\textbf{*yes}	&	\textbf{*yes	}&		&	1998-2008	\\
\cite{Bouet}	&		&		&	\emph{yes}	&		&		&	\emph{yes}	&	trade policies	&	2007-8 rise	\\
\cite{Rajagopal2009}	&	\emph{yes}	&	\emph{yes}	&	\textbf{*yes} (corn, soybeans) \textbf{*no} (rice and wheat)	&	\emph{yes}	&	\emph{yes}	&	\emph{yes}	&	trade policies, productivity decline	&	2007-8 rise	\\
\cite{Amponsah2009}	&	yes	&	yes	&	yes	&	yes	&	yes	&	yes	&		&	2007-8 rise \& fall	\\
\cite{Mondi2010}	&		&	no (rice)	&	maybe (rice)	&	yes (no rice)	&	no	&	yes	&		&	2007-8 rise \& fall	\\
\cite{Tyner2010_2}	&	yes	&	\emph{yes}	&	yes	&		&	yes	&	yes	&		&	2007-8 rise	\\
\cite{Ahrens2010}	&	yes	&	yes	&	yes	&	no	&	yes	&	yes	&	trade policies	&	2007-8 rise \& fall	\\
\cite{Searchinger2009_2}	&		&		&	yes	&		&		&		&		&	2007-8 rise	\\
\cite{Timilsina2010_2}	&		&		&	\textbf{*yes}	&		&		&		&		&	2007-8 rise	\\
\cite{Wright2010}	&	yes (long term)	&	\textbf{*no}	&	yes.	&		&	no	&		&	fertilizer cost: no; panic: yes	&	2007-8 rise \& fall	\\
\cite{Rutten2011}	&		&		&		&		&		&		&	trade policies	&	2007-8 rise \& fall, 2011 rise	\\
\cite{Rajcaniova2010}	&		&		&	maybe	&		&		&		&		&		\\
\cite{Meijerink2010}	&		&		&		&		&		&		&	trade policies	&		\\
\cite{Rajagopal2010}	&		&		&	yes	&		&		&		&		&	2007-8 rise	\\
\cite{Headey2009}	&	no	&	no	&	yes	&	maybe	&	yes	&		&	trade policies	&	2007-8 rise \& fall	\\
\cite{Simelton2010}	&		&		&		&		&		&		&	available land, climate change, population growth	&		\\
\cite{Abbott2010}	&		&		&		&	maybe	&		&		&		&		\\
\cite{Elliott2009}	&		&		&	yes	&		&		&		&		&	2007-8 rise \& fall	\\
\cite{Rismiller2009}	&		&		&	yes	&		&		&		&		&		\\
\cite{Singh2008}	&	\emph{yes}	&	\emph{yes}	&	yes	&	maybe	&	\emph{yes}	&	\emph{yes}	&		&	2007-8 rise \& fall	\\
\cite{Perry2009}	&		&		&	yes	&		&		&		&	delayed effect of easy credit	&	2007-8 rise	\\
\cite{Onour2010}	&		&		&		&		&	\textbf{*no}	&		&		&	1984-2009	\\
\cite{Clements2006}	&		&		&		&		&		&	\textbf{*yes} (2-5\%)	&		&	1975-2005	\\
\hline
	\end{tabular}
\caption{Literature review, part $2$. See text for notation.}
\label{tab:biblio2}
\end{sidewaystable}
\clearpage


\newpage
\phantomsection
\label{app:b}
\begin{center}
\Large Appendix B\\
\Large Commodity Prices: A Supply and Demand Model
\end{center}

In this appendix we present a simple model of commodity price formation based on supply and demand and show that, while the model is in general able to capture trends in prices before the year 2000, after this date other factors play a central role in determining  prices.  

Studies of supply-demand relationships for commodities have used two functional forms to characterize the price-quantity dependence \cite{ref:Spanos_2008}: a linear (constant slope) form \cite{ref:Breimyer_1961} and a log-linear (constant elasticity) form \cite{ref:Hughes_2006,ref:Hassan_1975}. If quantity demanded ($Q_d$) for a given commodity is determined by its price ($P$), the linear relationship is written as:

\begin{equation}
	Q_d(t) = \alpha_d - \beta_dP(t)
	\label{eq:lin}
\end{equation}

\noindent while the log-linear relationship is written as:

\begin{equation}
	\ln{Q_d(t)} = \ln{\alpha_d}- {\beta_d}\ln{P(t)}
	\label{eq:loglin}
\end{equation}

\noindent Even though in empirical studies the choice of functional form is an important decision, it is generally assumed that there is no \emph{a priori} reason for selecting either one \cite{Kulshreshtha1979}. A more general approach is given by the Box-Cox transformation \cite{Box1964},

\begin{equation}
	Q_d(t, \lambda) = \alpha_d - \beta_dP(t, \lambda)
	\label{eq:boxcox}
\end{equation}

\noindent where $\lambda$ is the parameter of transformation, such that

\begin{equation}
	\displaystyle
	Q_d(t, \lambda) = \frac{Q_d(t)^{\lambda}-1}{\lambda}
\end{equation}

\begin{equation}
	\displaystyle
	P(t, \lambda) = \frac{P(t)^{\lambda}-1}{\lambda}
\end{equation}

\noindent As $\lambda \rightarrow 1$, Eq. \ref{eq:boxcox} becomes Eq. \ref{eq:lin}, while as $\lambda \rightarrow 0$, Eq. \ref{eq:boxcox} becomes Eq. \ref{eq:loglin}.
Similar equations hold for supply.

We assume that, at each time step, $P(t)$ changes due to either a supply shock or a demand shock. In order to identify what kind of shock occurs, we define the surplus, $S(t)$, as the difference between production and consumption of the commodity at time $t$. We assume there is a positive demand shock at time $t$ if the surplus $S(t-1)$ is negative (shortage), so that the intercept of the demand curve shifts according to:

\begin{equation}
	\displaystyle
	\alpha_d(t) = \alpha_d(t-1) - S(t-1)
\end{equation}

\begin{figure}[t]
\refstepcounter{figref}\label{fig:fits}
\begin{center}$
\begin{array}{cccc}
\href{http://necsi.edu/research/social/img/fig5a.pdf}{\includegraphics[width=0.5\linewidth, trim= 0cm 1.8cm 1.3cm 0.3cm, clip]{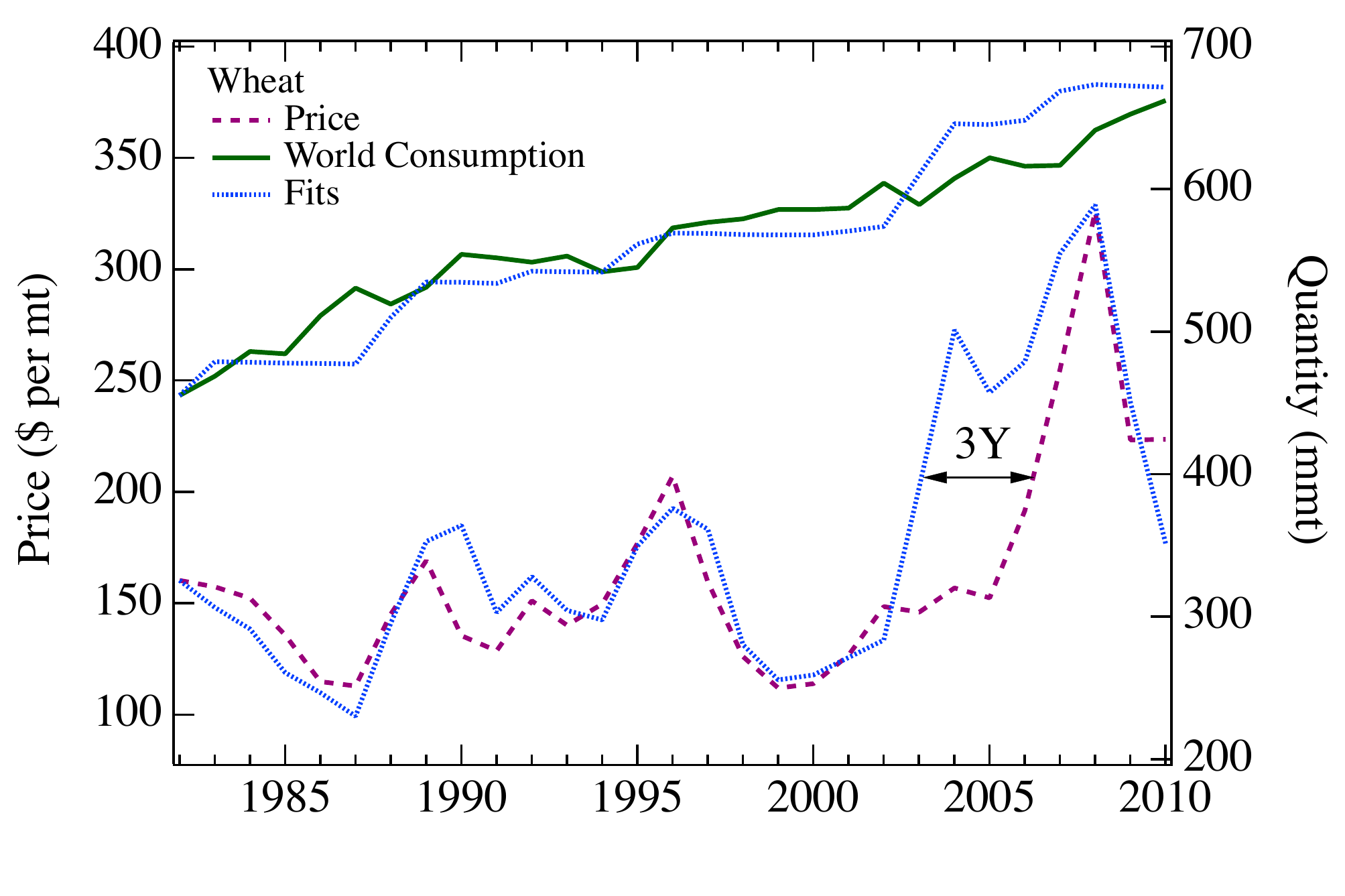}} &
\href{http://necsi.edu/research/social/img/fig5b.pdf}{\includegraphics[width= 0.5\linewidth, trim= 1.3cm 1.8cm 0cm 0.3cm, clip]{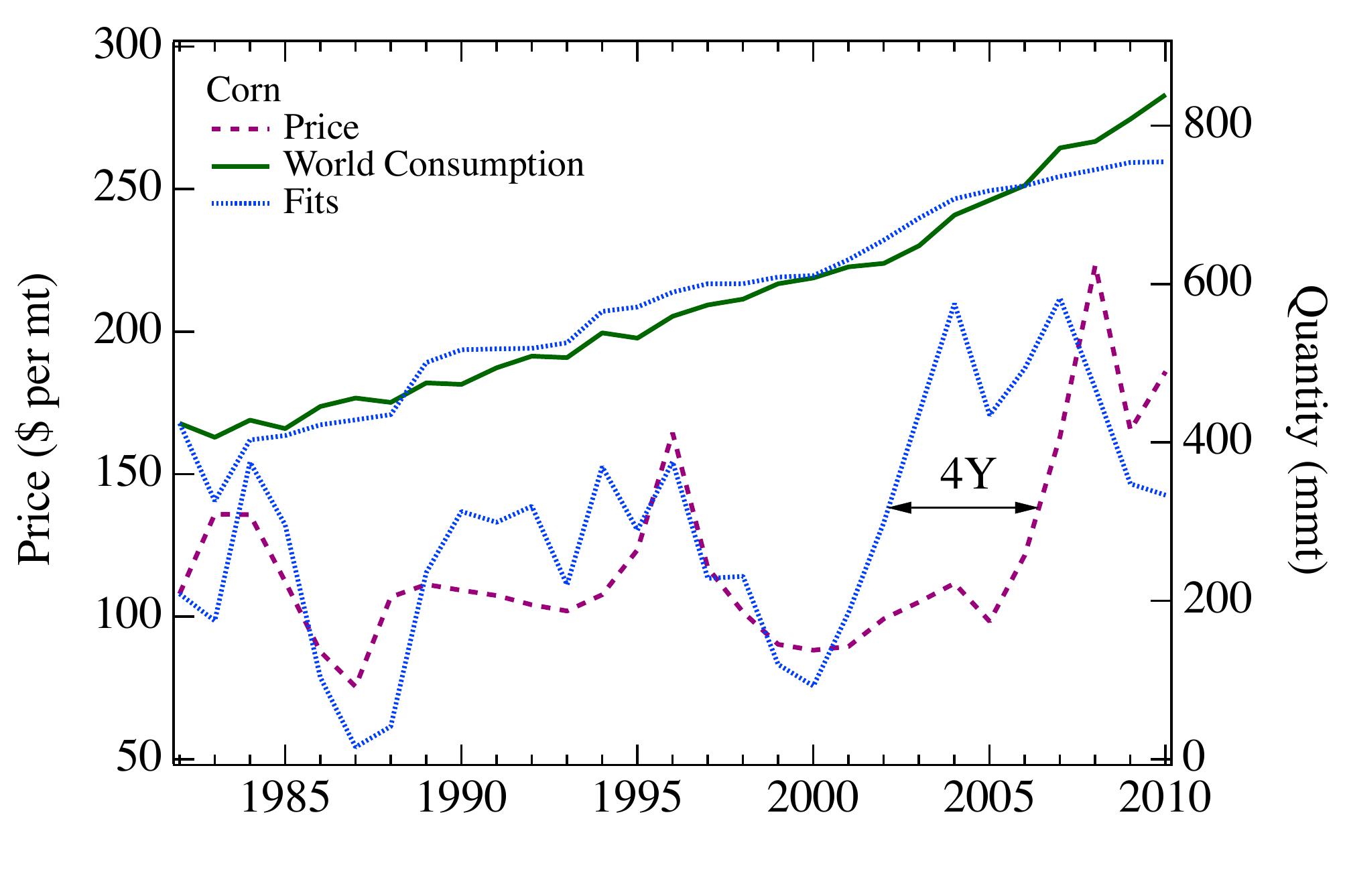}} \\
\href{http://necsi.edu/research/social/img/fig5c.pdf}{\includegraphics[width= 0.5\linewidth, trim= 0cm 1cm 1.3cm 0.3cm, clip]{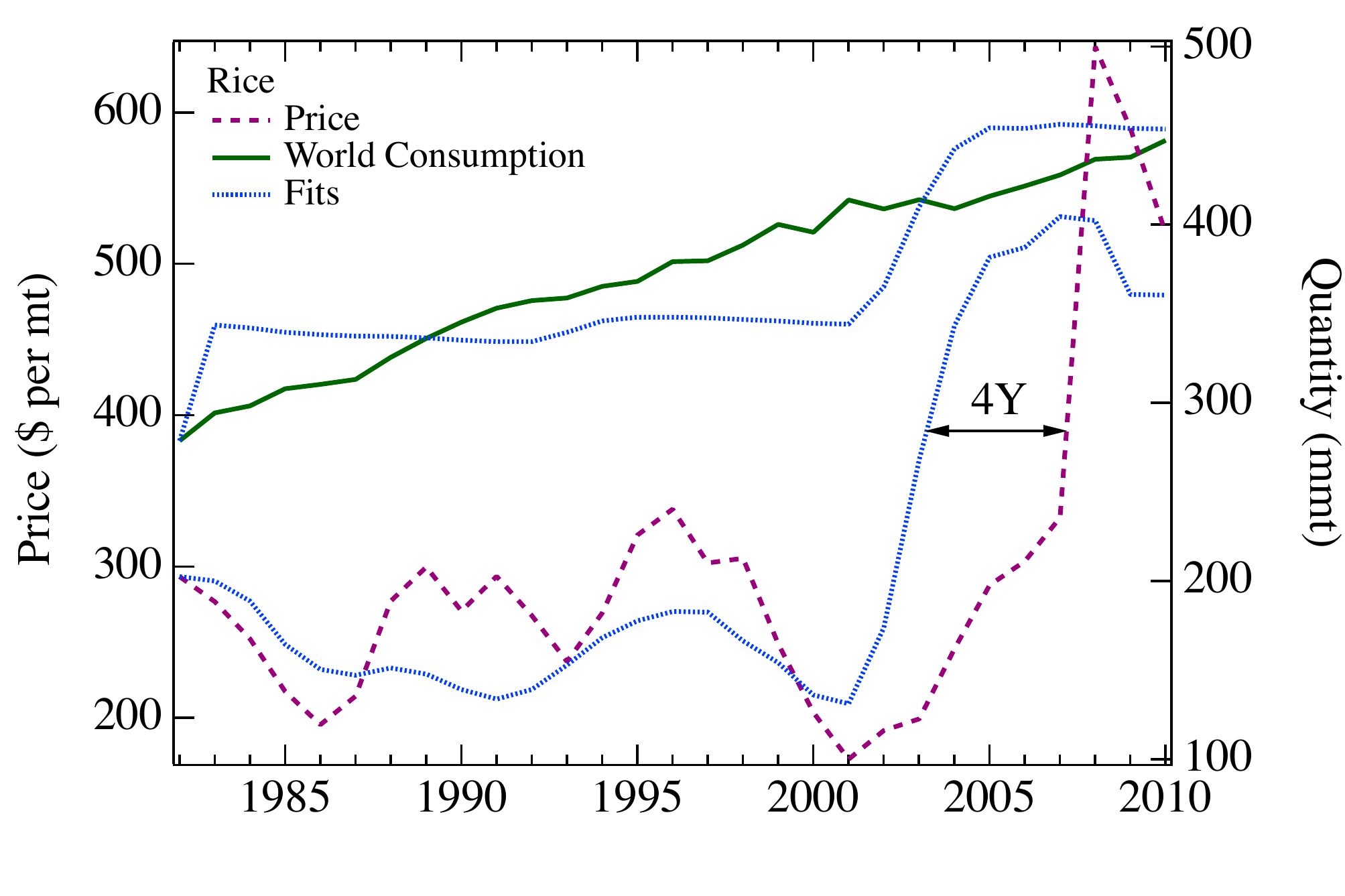}} &
\href{http://necsi.edu/research/social/img/fig5d.pdf}{\includegraphics[width= 0.5\linewidth, trim= 1.3cm 1cm 0cm 0.3cm, clip]{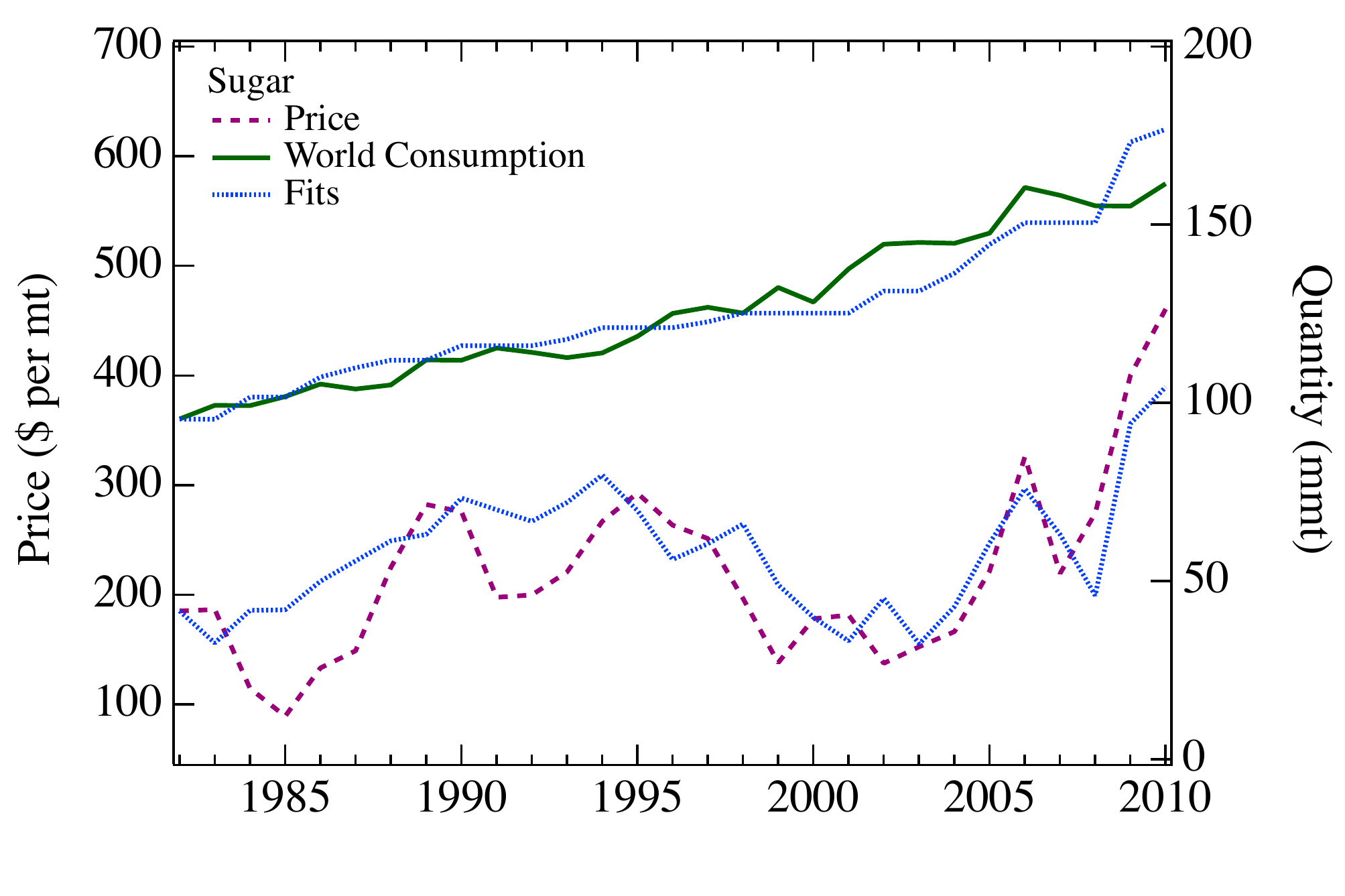}}\\
\end{array}$
\end{center}
\caption{\textbf{Supply and demand model} - For wheat, corn, rice, sugar (left to right and top to bottom). Price time series of the commodity (dashed purple lines) and global consumption time series (solid green lines). Blue dotted curves are best fits according to Eq. \ref{eq:pe} and Eq. \ref{eq:qe} for prices and consumption, respectively. Annual supply-demand values are from \cite{source_grains}, and prices from \cite{source_oil}. Values of the fitting parameters: wheat - $\lambda = 1\pm0.01$, $\alpha_d(1982) = 309\pm53$ and $\beta_d=1.01\pm0.13$, corn - $\lambda = 0.94\pm0.01$, $\alpha_d(1982) = 114\pm22$ and $\beta_d=1.91\pm0.23$, rice - $\lambda =0.95\pm0.01$, $\alpha_d(1982) = 87\pm77$ and $\beta_d=0.64\pm0.19$, sugar - $\lambda = 0.86\pm0.07$, $\alpha_d(1982) = 26\pm20$ and $\beta_d=0.31\pm0.12$.}
\end{figure}

\noindent An analogous argument can be made for a supply shock, with appropriate signs. At each time step we can therefore estimate both price and quantity at equilibrium:

\begin{equation}
	\displaystyle
	P_e(t, \lambda)=\frac{\alpha_d(t)-\alpha_s(t)}{\beta_d+\beta_s}
	\label{eq:pe}
\end{equation}
\begin{equation}
	\displaystyle
	Q_e(t,\lambda)=\frac{\alpha_d(t)\beta_s+\alpha_s(t)\beta_d}{\beta_d+\beta_s}
	\label{eq:qe}
\end{equation}

\noindent where $\alpha_s(t)$ and $\beta_s$ are the intercept and slope of the supply curve, respectively. We use Eq. \ref{eq:pe} to fit the price time series of a given commodity $P(t)$, and Eq. \ref{eq:qe} to fit its consumption, $Q_d(t)$. The only input into the model is the surplus $S(t)$, and three parameters are used for fitting the price and consumption: the transformation parameter $\lambda$, one slope (either $\beta_d$ or $\beta_s$) and the initial value of one of the intercepts (either $\alpha_d(0)$ or $\alpha_s(0)$). The other slope/intercept is determined by setting initial values to empirical data, $P_e(0)=P(0)$ and $Q_e(0)=Q(0)$. Empirical price data was adjusted for the US consumer price index, so that $P_e(t, \lambda)$ represents constant prices.

Fig. \ref{fig:fits} shows that this simple model is able to capture most features of commodity price fluctuations for wheat, corn, rice and sugar, but it fails to reproduce the 2006-2008 spike. Instead, the model predicts price peaks starting from 2000-2002. However, in order to fit the spike in the price time series, the model creates a jump in demand which does not occur in the actual data, as demonstrated by the poor fit of the consumption time series.
Absent a mechanism for shifting of price increases by a time delay of 3-4 years, distinct causes of the supply-demand change in 2000-2002 and price peaks of 2006-2008 are necessary. Policy-based reductions in reserves are a possible explanation of the changes in 2000-2002 \cite{Huang2008, Abbott2009}.  Appendix \hyperref[app:d]{D} demonstrates that commodity speculation can account for the peaks in 2007-8.


\newpage
\phantomsection
\label{app:c}
\begin{center}
\Large Appendix C\\
\Large Corn Ethanol and Food Prices
\end{center}

Currently, the two main uses of corn are livestock feed (consuming 41.4\% of the US supply in 2010), and ethanol production (40.1\%) \cite{ref:wasde_2011}. Other uses include direct human consumption and the production of oil, sweeteners and starch for use in a wide range of processed foods. Corn is therefore heavily used as an input for many food sectors, and it is reasonable to consider the amount of corn used to produce ethanol to have an impact on food prices \cite{ref:Fortenbery_2008}. The proportion of corn for ethanol has increased from 6\% to its current value of 40\% over the last 10 years.

We begin by assuming a linear dependence of the corn quantity supplied for food use, $Q_f(t)$, on the price of food, $P(t)$, as in the model described in Appendix \hyperref[app:b]{B} with $\lambda =1$, leading to the equilibrium equations for price and quantity:

\begin{equation}
	\displaystyle
	P(t)=\frac{\alpha_d(t)-\alpha_s(t)}{\beta_d+\beta_s}
	\label{eq:pe_corn}
\end{equation}
\begin{equation}
	\displaystyle
	Q_f(t)=\frac{\alpha_d(t)\beta_s+\alpha_s(t)\beta_d}{\beta_d+\beta_s}
	\label{eq:qe_corn}
\end{equation}
We now assume that the use of corn for ethanol production, $Q_x(t)$, causes a dominant supply shock for corn used directly and indirectly for food, so that

\begin{equation}
	\displaystyle
	\alpha_s(t) \sim Q_f(t) \sim Q_t(t)-Q_x(t)
\end{equation}

\noindent where $Q_t(t)$ is the total amount of corn produced. In this model, a change in price would then be caused only by supply shocks. This does not imply that the supply and demand aside from corn ethanol production is static. For example, a growing world population creates a growing demand, but if this demand is met by a corresponding growing supply prices need not change. The total quantity demanded at equilibrium would then follow the shifts of the demand intercept, so that the difference $Q_t(t)-\alpha_d(t)\approx Q_t(t_0)-\alpha_d(t_0)$ would not be time dependent. The assumption that ethanol is a dominant shock is equivalent to the assumption that the price without the ethanol production would be relatively constant. Substituting in Eq. \ref{eq:pe_corn} and dropping the time dependence for the total corn production and the demand intercept yields

\begin{figure}[t]
\refstepcounter{figref}\label{fig:fit_corn}
\begin{center}
\href{http://necsi.edu/research/social/img/fig6.pdf}{\includegraphics[width=1\linewidth, trim= 0cm 1cm 0cm 0cm, clip]{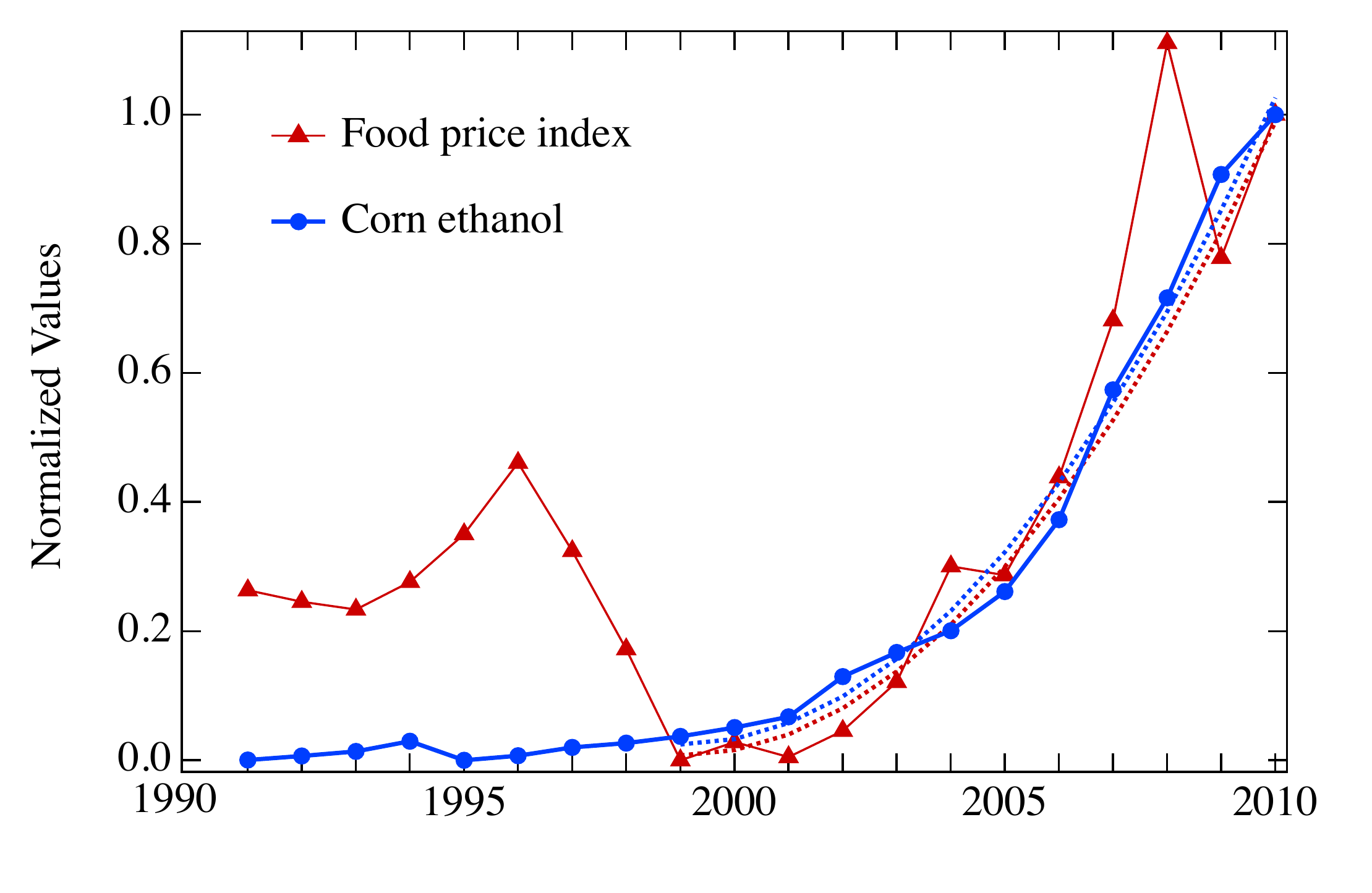}}
\end{center}
\caption{\textbf{Model results} - Annual corn used for ethanol production in the US (blue circles) and the FAO Food Price Index from 1991-2010 (red triangles). Values are normalized to range from 0 (minimum) to 1 (maximum) during this period. Dotted lines are best fits to quadratic growth, with quadratic coefficients of $0.0083 \pm 0.0003$ for corn ethanol and $0.0081 \pm 0.0003$ for FAO index. Goodness of fit is measured with the coefficient of determination, $R^2=0.989$ for corn and $R^2=0.986$ for food. The 2007-2008 peak was not included in the fit or the normalization of the FAO index time series.}
\end{figure}

\begin{equation}
	\displaystyle
	Q_x(t)=(\beta_d+\beta_s)P(t)+Q_t-\alpha_d
	\label{eq:qx_corn}
\end{equation}
The supply and demand model of Eq.  \ref{eq:qx_corn} can be considered a quite general first order model of the food index as a function of many factors, $P = P(f_i)$, one of which is the amount of corn to ethanol conversion. We can perform a Taylor series expansion of such a generalized function with respect to its dependence on ethanol due to both direct and indirect effects. Even though the change in corn ethanol production is a large fraction of US corn production, we can consider an expansion to linear order of its effect on food prices generally. The total change in price, $\Delta P$, is then written as a sum over partial derivative relative to all factors and their change with respect to corn ethanol production:
\begin{equation}
	\displaystyle
	\Delta P=\left(\frac{d P}{d Q_x}\right) Q_x = \sum_i \left(\frac{\partial P}{\partial f_i}\right) \left(\frac{\partial f_i}{\partial Q_x}\right) Q_x
\end{equation}
where the reference price is for $Q_x=0$. Comparing with the previous equation we see that this has the same behavior as a supply and demand model with an effective elasticity given by:
\begin{equation}
	\displaystyle
	\beta_d+\beta_s = \left(\sum_i \left(\frac{\partial P}{\partial f_i}\right) \left(\frac{\partial f_i}{\partial Q_x}\right) \right)^{-1}
\end{equation}
The approximation that is needed for validity of this expression is that the dominant shock in the agriculture and food system is due to the corn use for ethanol. The validity of this assumption may be enhanced by the cancellation of other effects that contribute to both increases and decreases in prices.  

Thus, if the assumptions of the model are correct, the change in quantity of corn ethanol would be proportional to the change in food price. This implies that the time dependence of the corn used for ethanol and of the food price index should each have the same functional form. The existence of an ethanol production byproduct use for feed (distillers grain), which is a fixed proportion of the corn, does not influence the functional form. We test this hypothesis in Fig. \ref{fig:fit_corn}, where we plot both the time dependence of corn ethanol and the FAO food price index between 1999 and 2010. Values are normalized to range from 0 (minimum) to 1 (maximum) during this period in order to compare the functional forms of the two curves. If we exclude the 2007-2008 price peak, both curves can be accurately fit by quadratic growth ($R^2$ values of $0.986$ for food prices and $0.989$ for ethanol fraction). The Pearson correlation of the two curves is $\rho=0.98$.

This model differs from the supply-demand model described in Appendix \hyperref[app:b]{B} in that here we consider the total value of the change in ethanol use (i.e. $Q_x(t)$), and not just the surplus as reflected in reserves. The combination of the large change in food prices and the large change in the amount of grain used for ethanol production (over 15\% of total global corn production), along with the proportionality we find between the two quantities, is strong evidence for a causal link between them.


\newpage
\phantomsection
\label{app:d}
\begin{center}
\Large Appendix D\\
\Large A Dynamic Model of Speculators
\end{center}

In this appendix we present a simple dynamic model of the role of trend-following speculators and their ability to cause deviations from equilibrium supply and demand prices. In Appendix \hyperref[app:e]{E} we will augment the model to incorporate the specific conditions of the commodity markets, the demand shock of corn to ethanol conversion from Appendix \hyperref[app:c]{C}, and investor shifting between markets.

Our model directly describes the possibility of speculators causing price deviations from equilibrium supply and demand. A  progressive departure from equilibrium leads to supply and demand conditions increasingly countering that deviation. The interplay of these effects leads to the oscillations of bubble and crash dynamics. 

\begin{figure}[t]
\refstepcounter{figref}\label{fig:nospec}
\center
$
\begin{array}{cccc}
\href{http://necsi.edu/research/social/img/fig7a.pdf}{\includegraphics[width=0.5\linewidth]{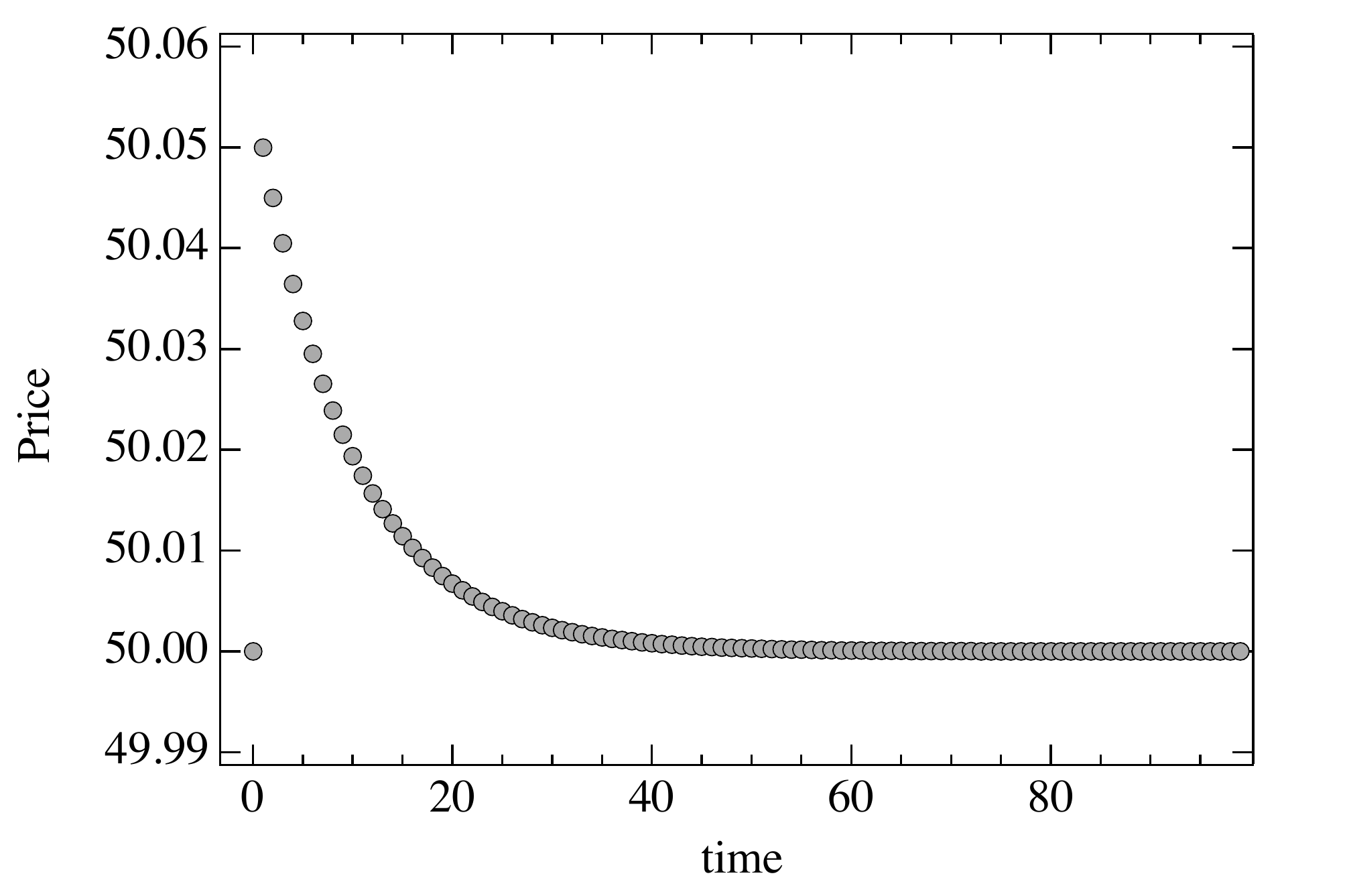}}&
\href{http://necsi.edu/research/social/img/fig7b.pdf}{\includegraphics[width=0.5\linewidth]{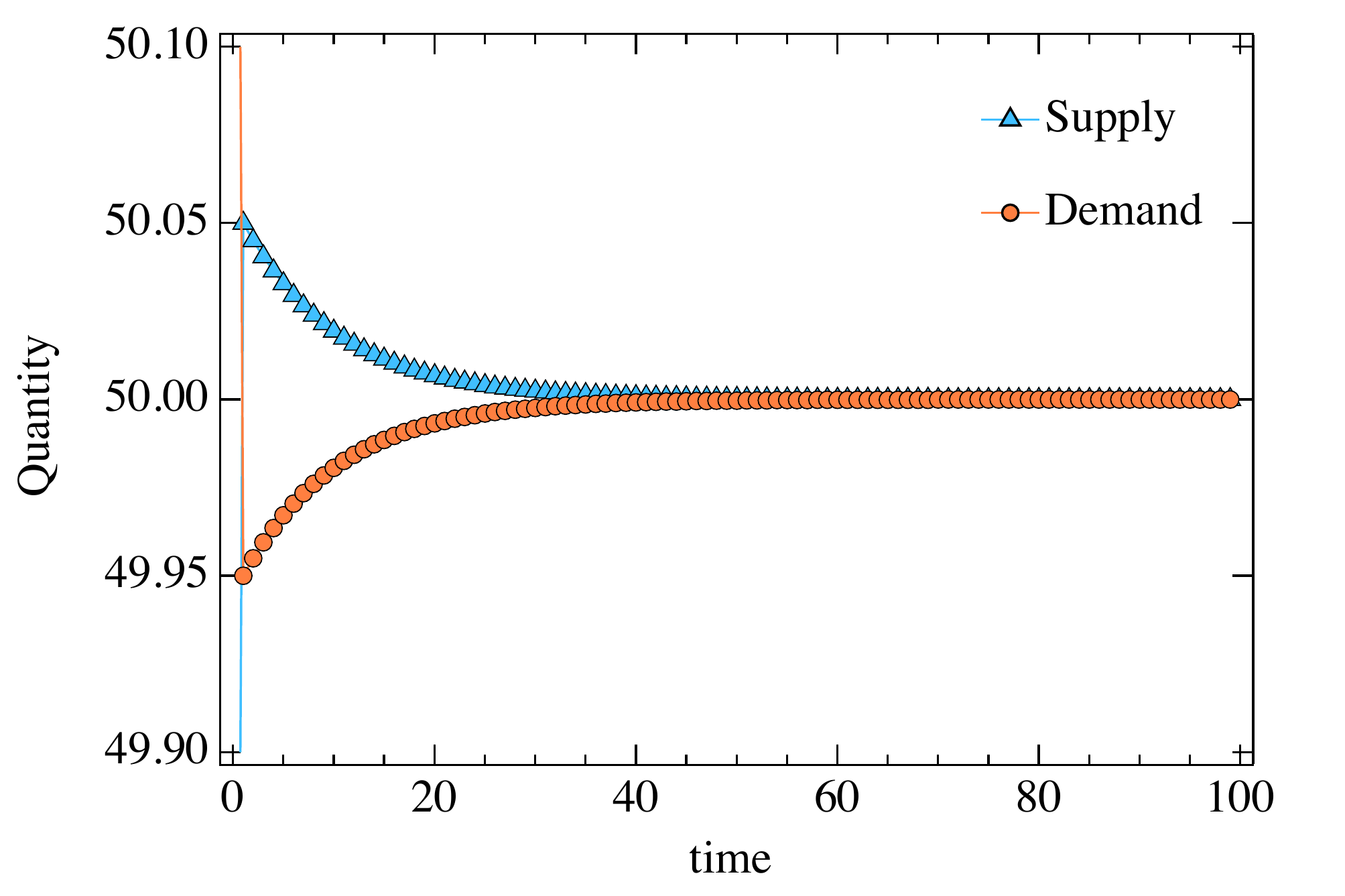}}\\
\href{http://necsi.edu/research/social/img/fig7c.pdf}{\includegraphics[width=0.5\linewidth]{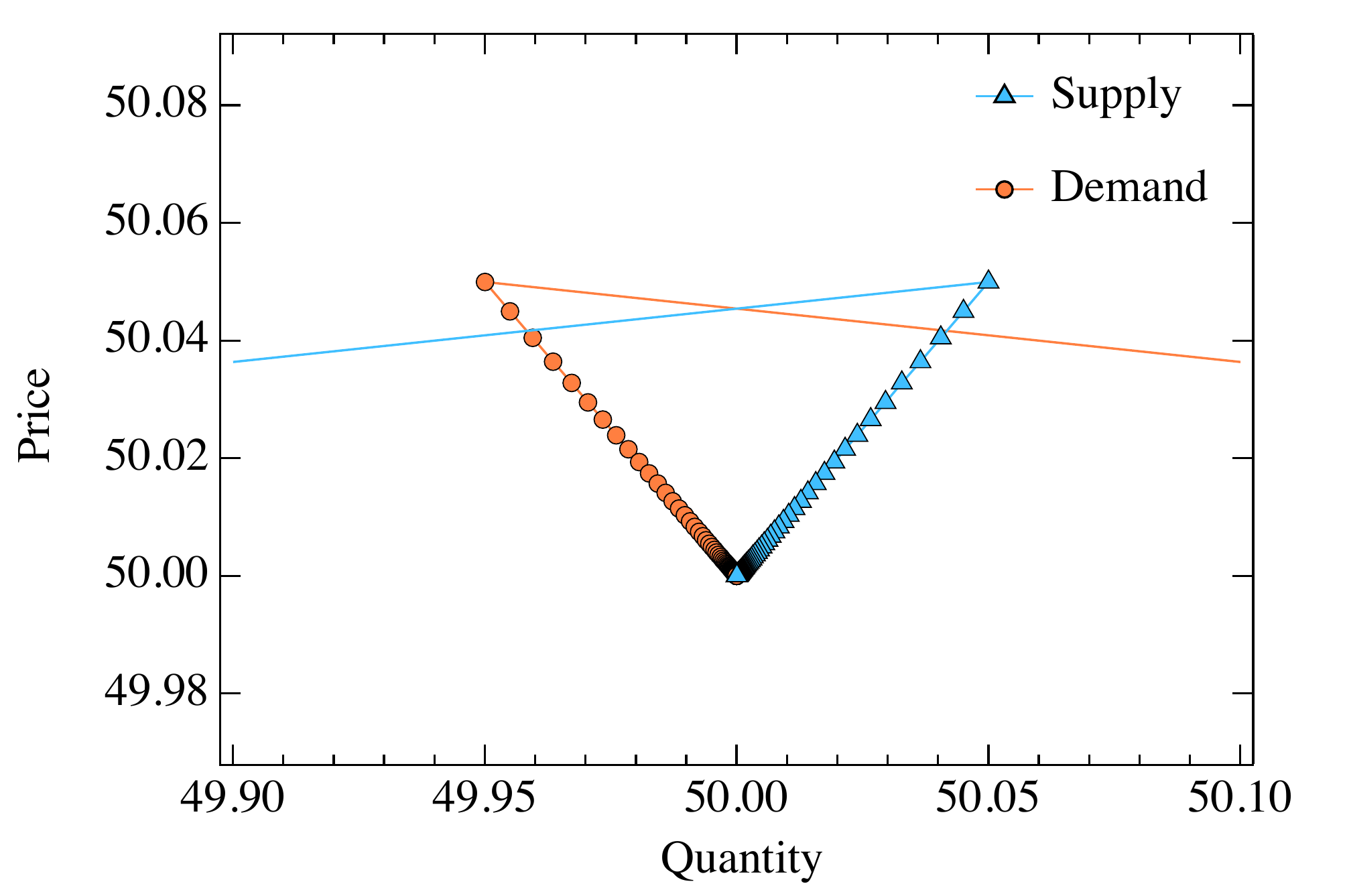}}&
\href{http://necsi.edu/research/social/img/fig7d.pdf}{\includegraphics[width=0.5\linewidth]{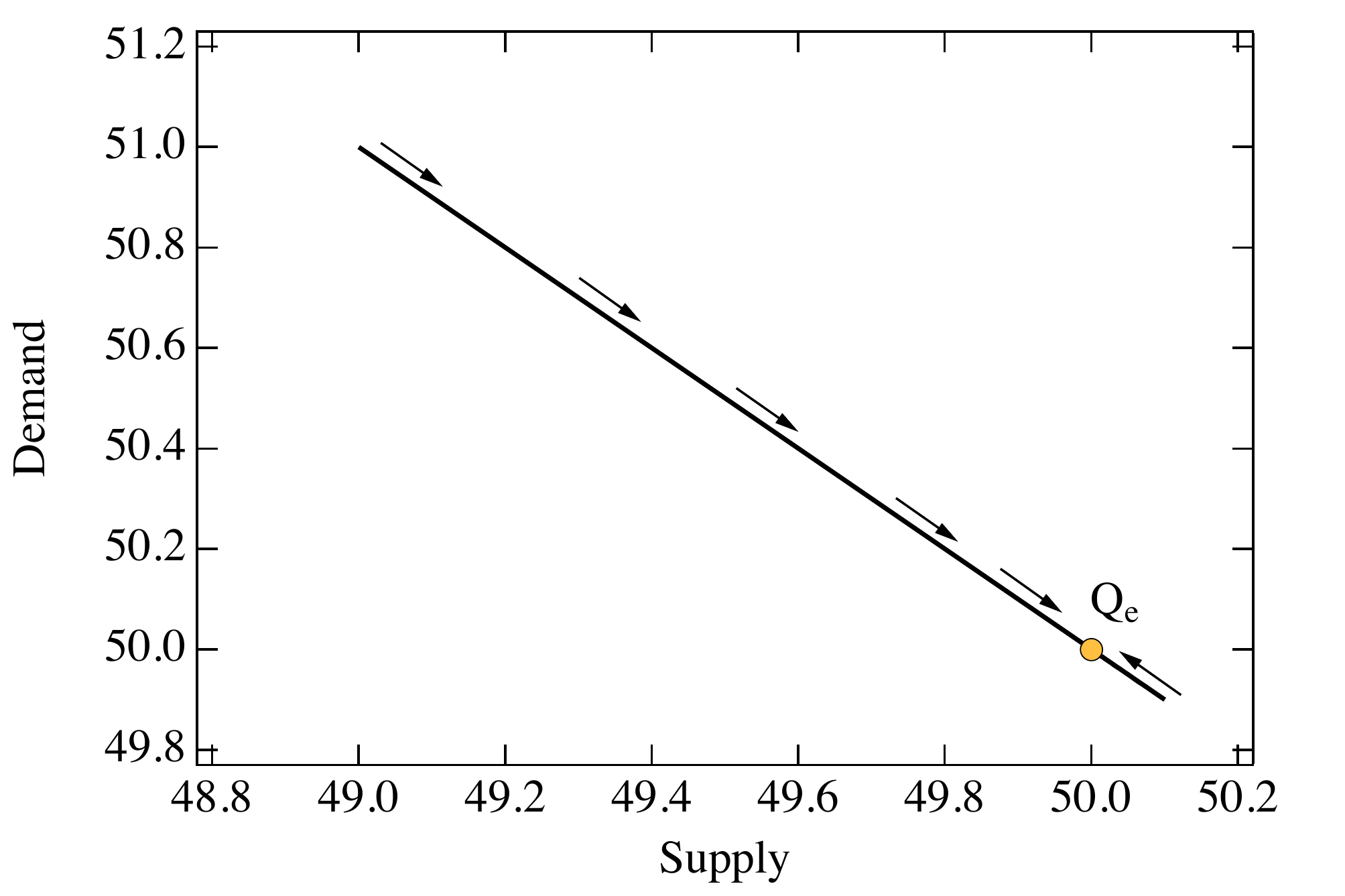}}
\end{array}
$
\caption{\textbf{Dynamics without speculators} - Dynamic response of the system to a supply/demand shock at time $t=0$. \emph{Top Left}: Exponential convergence of $P$ to its equilibrium value. \emph{Top Right}: Supply and demand as a function of time. \emph{Bottom Left}: Price/quantity relationship. \emph{Bottom Right}: Dynamic evolution of $Q_d$ vs $Q_s$. Value of parameters: $k_{sd}=0.1$, $k_c=5$.}
\end{figure}

We construct our speculator model starting from a supply and demand one. The price dynamics based upon supply and demand for a single commodity can be represented by \cite{ref:Ferguson_1998}:

\begin{equation}
Q_d(t)=\alpha_d-\beta_dP(t)
\end{equation}
\begin{equation}
Q_s(t)=\alpha_s+\beta_sP(t)
\end{equation}
\begin{equation}
P(t+1)=P(t)+\gamma_0(Q_d(t)-Q_s(t))
\label{eq:walras}
\end{equation}
where $Q_d (t)$ is the quantity demanded at time $t$, $Q_s(t)$ is the quantity supplied and $P(t)$ is the price of the commodity. We assumed a linear relationship between quantity and price, and we replaced the equilibrium condition, $Q_e=Q_d=Q_s$, with Eq. \ref{eq:walras}, the Walrasian adjustment mechanism \cite{ref:McDonald_1980}: $P$ rises if the demand exceeds supply and vice versa, where $\gamma_0$ is the strength of the restoring force toward equilibrium.
This is equivalent to a single first order difference equation in $P$ \cite{ref:Hamilton_1994},

\begin{equation}
P(t+1)+P(t)(k_{sd}-1)=k_c,
\label{eq:nospec} 
\end{equation}
where $k_{sd}=\gamma_0(\beta_d+\beta_s)$ and $k_{c}=\gamma_0(\alpha_d-\alpha_s)$. This can be solved to give:

\begin{equation}
P(t)=(P_1-P_e)(1-k_{sd})^t+P_e
\label{eq:cob}
\end{equation}
where $(P_1-P_e)$ is the initial deviation from the equilibrium and $P_e=P_0=k_c/k_{sd}$ the equilibrium price. This behavior is summarized in Figure \ref{fig:nospec}, where $k_{sd}<1$ and $Q_d$ and $Q_s$ are displaced by a small percentage from their equilibrium value at $t=0$, in order to simulate a supply/demand shock. Eq. \ref{eq:cob} is similar to the solution of the classic Cobweb Model \cite{ref:Kaldor_1938}, with the difference that the term raised to the $t_{th}$ power is proportional to the ratio of the supply and demand slopes, $1-k_{sd}=-\beta_s/\beta_d$, and not to their sum. Therefore, as in the Cobweb Model, we can have convergence ($k_{sd}<2$) or divergence ($k_{sd}>2$) depending on the slope of the linear response of supply and demand to prices.

We now introduce the influence of trend-following speculators. If the price change of the commodity is positive in the previous time step, speculators are willing to buy a quantity $\mu[P(t)-P(t-1)]$ of commodity, otherwise they sell $\mu[P(t-1)-P(t)]$. 
The quantity bought (sold) is added (subtracted) from the term $(Q_d(t)-Q_s(t))$ in price setting Eq.~\ref{eq:walras}. 
The result is a non-homogeneous second order difference equation in $P$ of the type $aP(t+1)+bP(t)+cP(t-1)=g$, so that Eq. \ref{eq:nospec} becomes:

\begin{equation}
P(t+1)+P(t)(k_{sd}-k_{sp}-1)+P(t-1)k_{sp}=k_c
\label{eq:spec} 
\end{equation}
where $k_{sp}=\mu\gamma_0$. The values of the coefficients are given in terms of both the supply and demand parameters and the speculator response parameter $\mu$. If prices are measured in units of $k_c$, Eq. \ref{eq:spec} can be normalized to:

\begin{equation}
p(t+1)+p(t)(k_{sd}-k_{sp}-1)+p(t-1)k_{sp}=1
\label{eq:spec2} 
\end{equation}
where $p(t)=P(t)/k_c$.

\begin{figure}[t]
\refstepcounter{figref}\label{fig:spec_095}
\center
$
\begin{array}{cccc}
\href{http://necsi.edu/research/social/img/fig8a.pdf}{\includegraphics[width=0.5\linewidth]{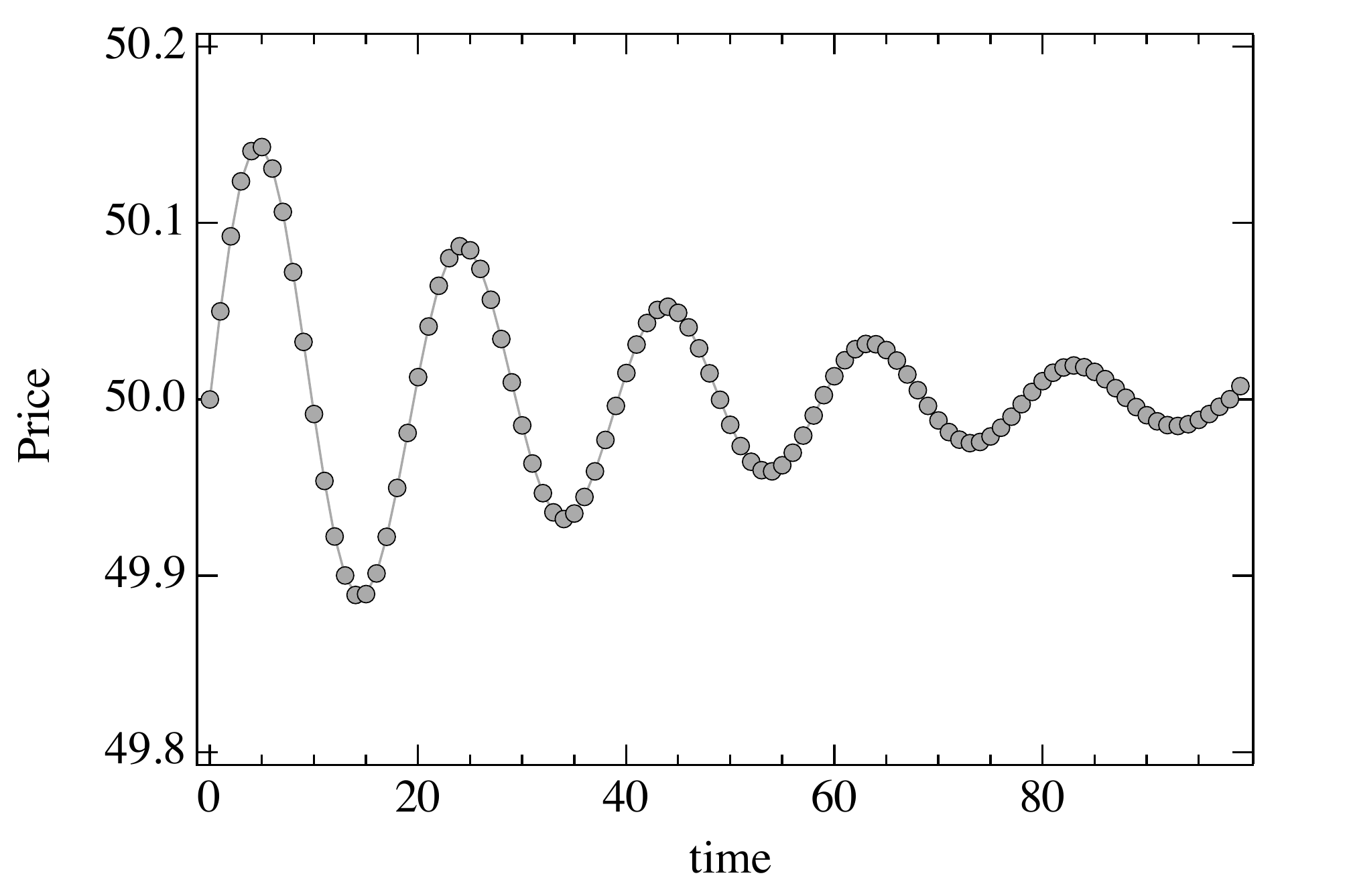}}&
\href{http://necsi.edu/research/social/img/fig8b.pdf}{\includegraphics[width=0.5\linewidth]{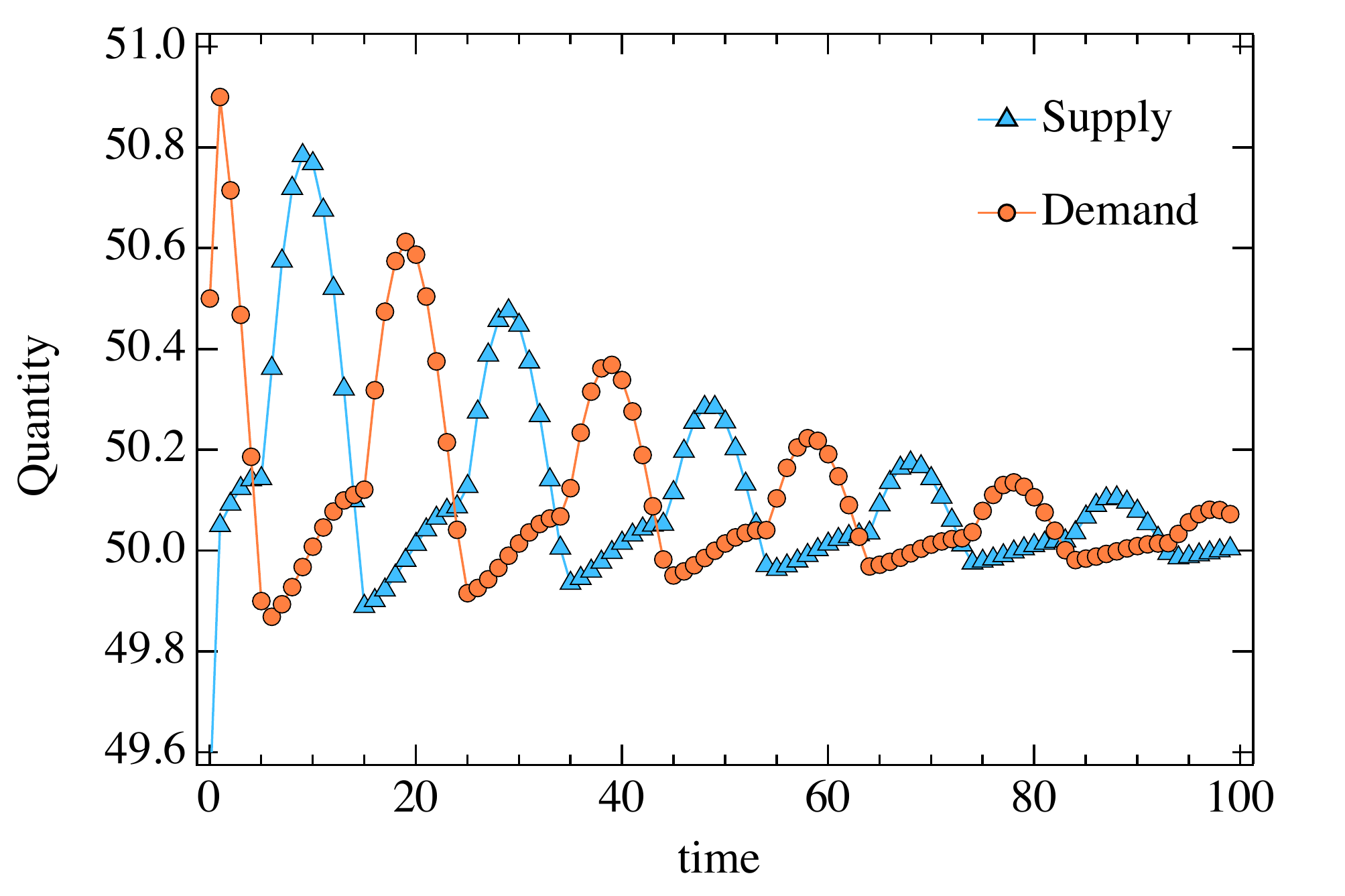}}\\
\href{http://necsi.edu/research/social/img/fig8c.pdf}{\includegraphics[width=0.5\linewidth]{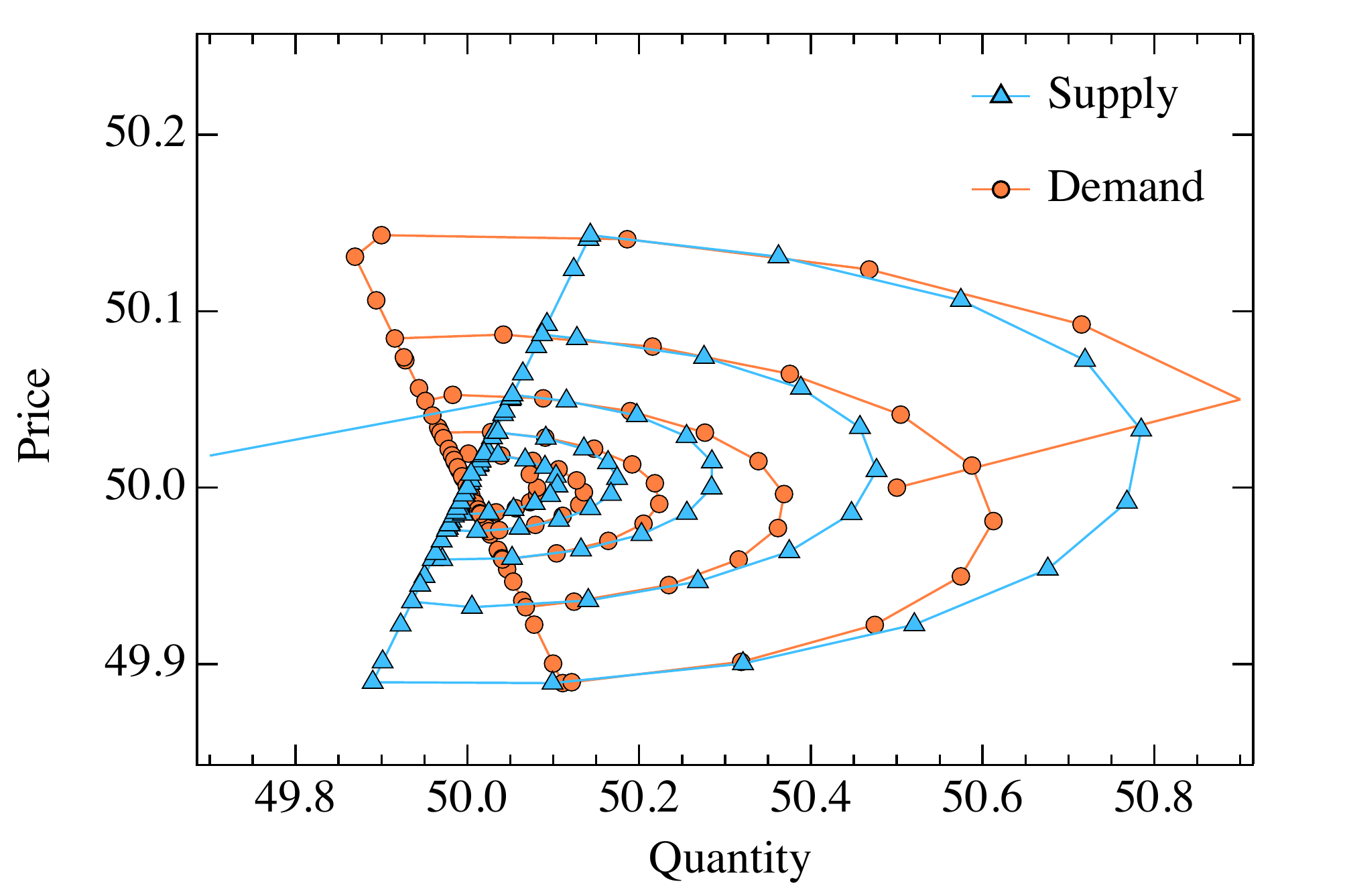}}&
\href{http://necsi.edu/research/social/img/fig8d.pdf}{\includegraphics[width=0.5\linewidth]{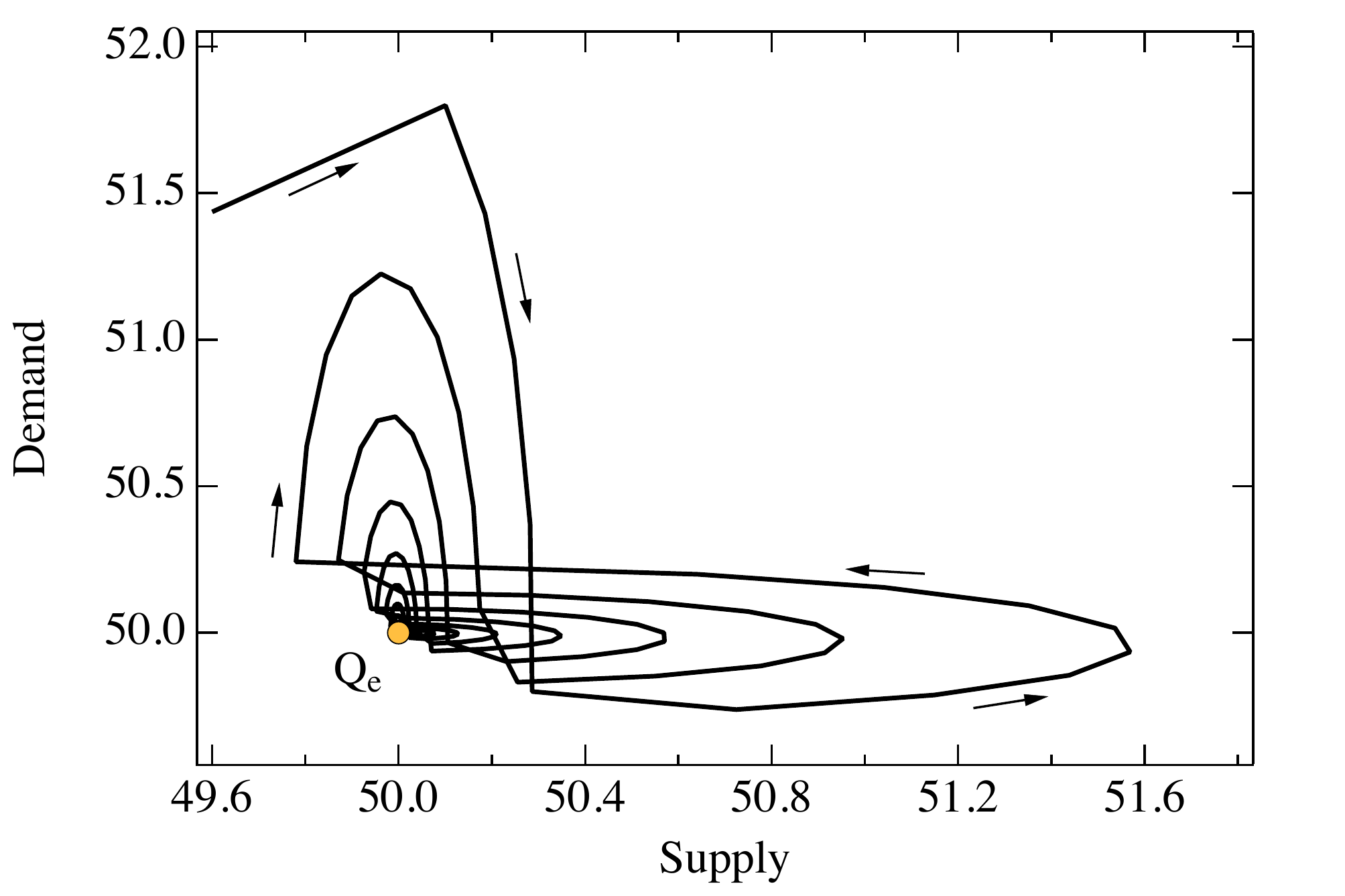}}
\end{array}
$
\caption{\textbf{Dynamics with speculators: $k_{sp}<1$} - Dynamic response of the system to a supply/demand shock at time $t=0$. \emph{Top Left}: Oscillating convergence of $P$ to its equilibrium value. \emph{Top Right}: Supply and demand as a function of time. \emph{Bottom Left}: Price/quantity relationship. \emph{Bottom Right}: Dynamic evolution of $Q_d$ vs $Q_s$ towards $Q_e$.}
\end{figure}

Three cases have to be considered, depending on the value of the discriminant $\delta=b^2-4ac$. We again consider a small displacement from equilibrium at $t=0$, so that $P_0=P_e$.\\

\textbf{Case 1: $\delta>0$}

If the discriminant is positive, there are two distinct roots for the characteristic equation of a second order difference equation, and Eq. \ref{eq:spec} can be solved to give:

\begin{equation}
P(t)=\delta^{-1/2}(P_1-P_e)(m_1^t-m_2^t)+P_e
\label{eq:case1}
\end{equation}

\begin{figure}[t]
\refstepcounter{figref}\label{fig:spec_1}
\center
$
\begin{array}{cccc}
\href{http://necsi.edu/research/social/img/fig9a.pdf}{\includegraphics[width=0.5\linewidth]{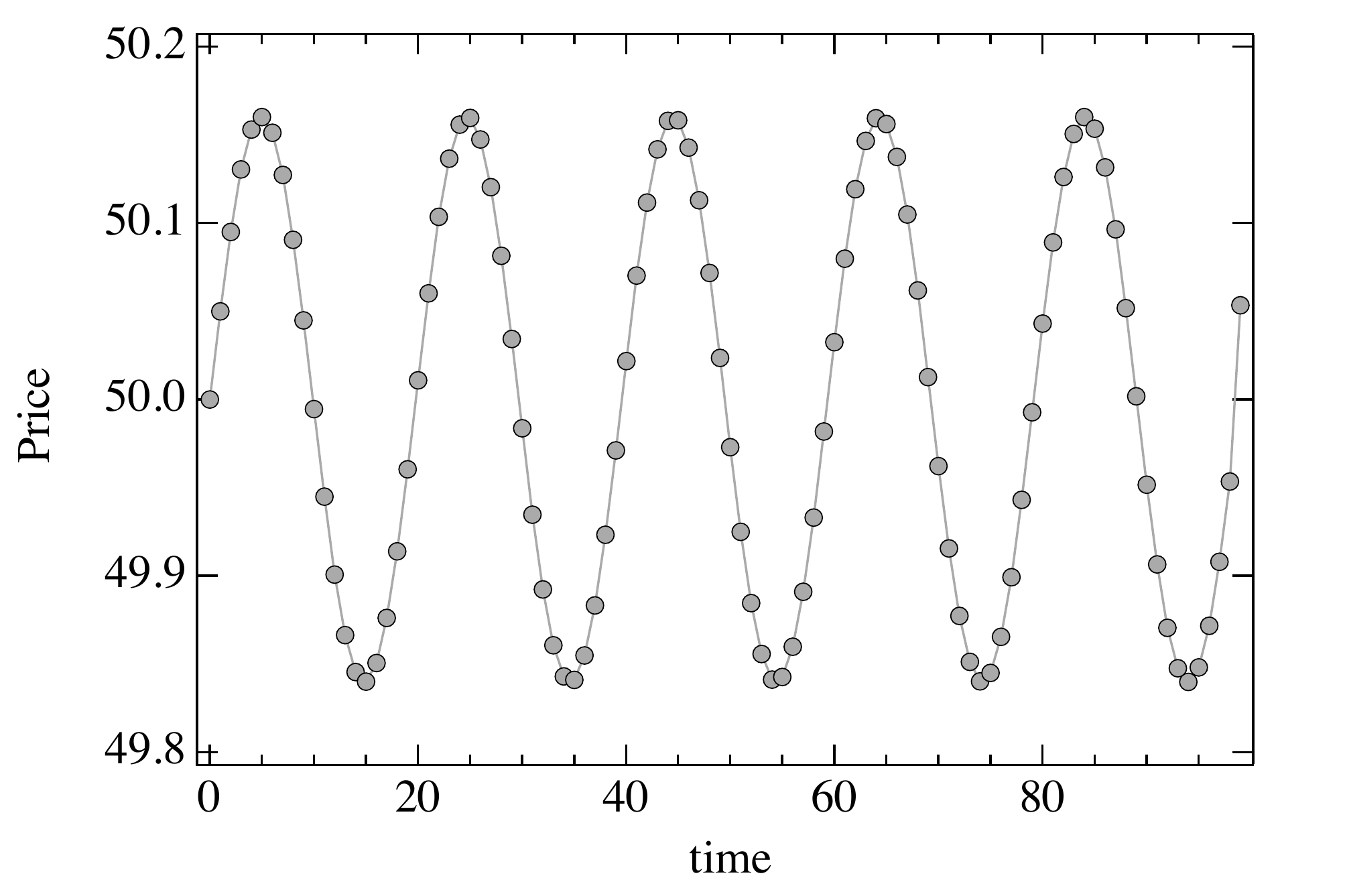}}&
\href{http://necsi.edu/research/social/img/fig9b.pdf}{\includegraphics[width=0.5\linewidth]{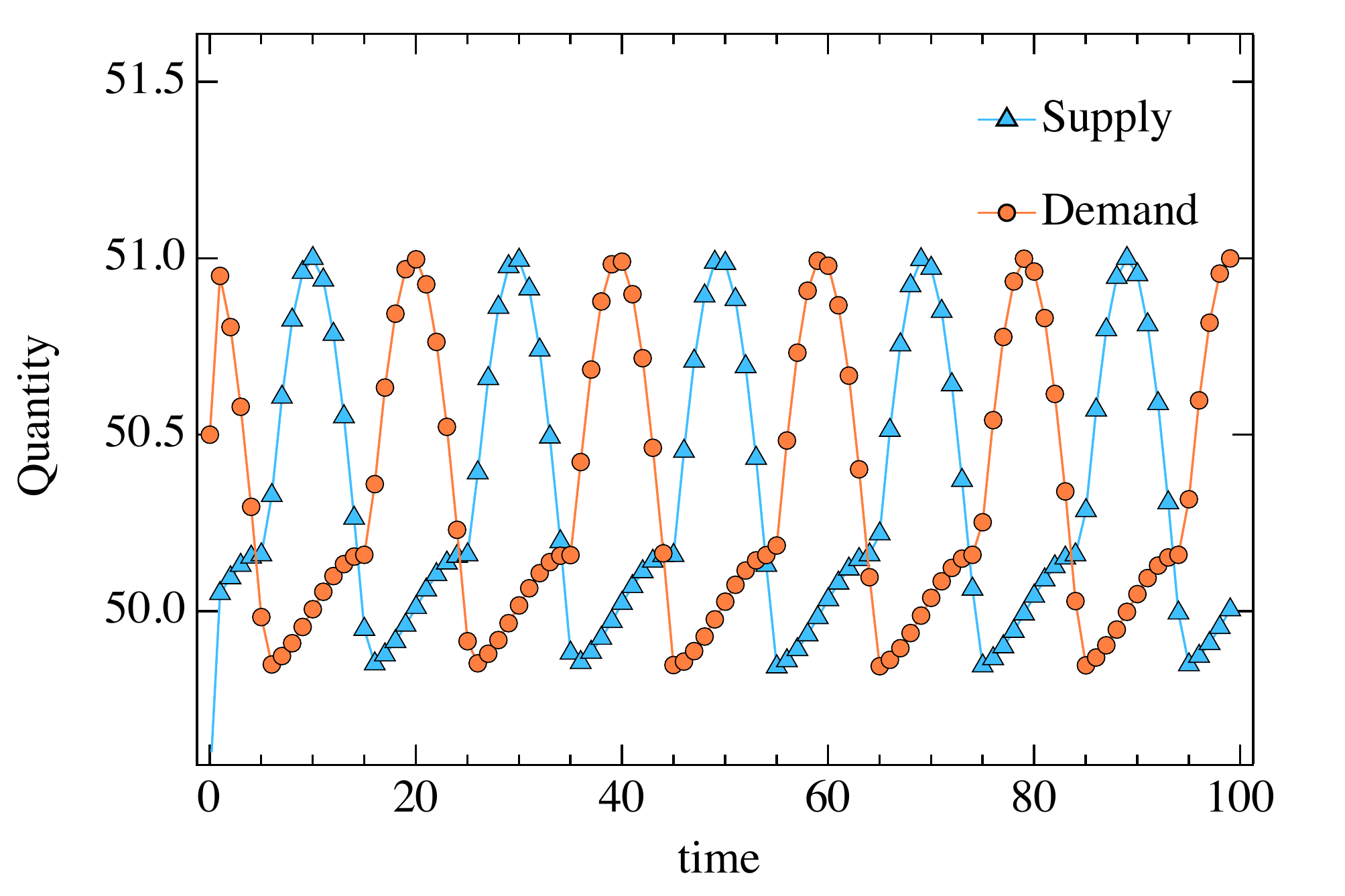}}\\
\href{http://necsi.edu/research/social/img/fig9c.pdf}{\includegraphics[width=0.5\linewidth]{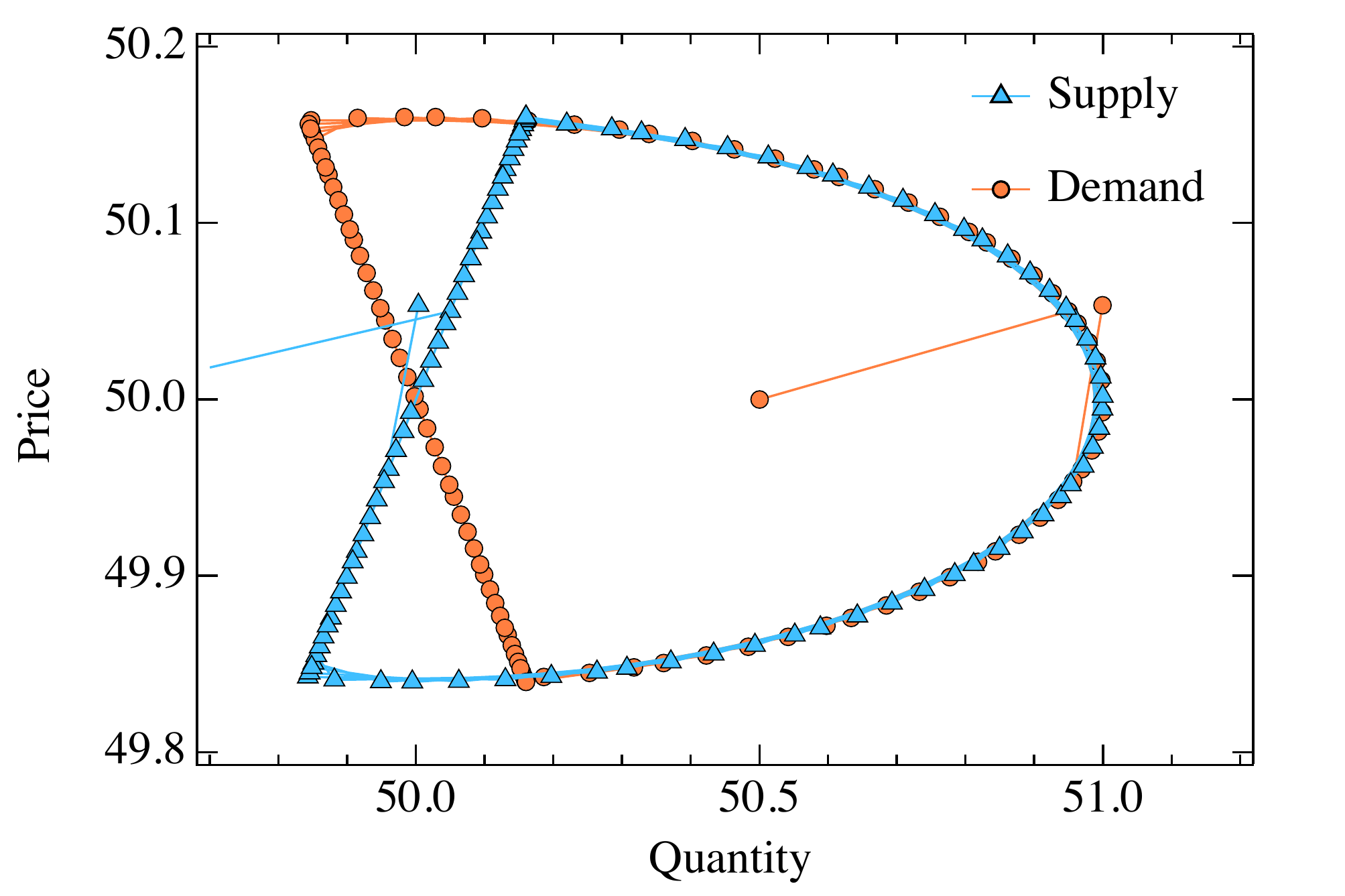}}&
\href{http://necsi.edu/research/social/img/fig9d.pdf}{\includegraphics[width=0.5\linewidth]{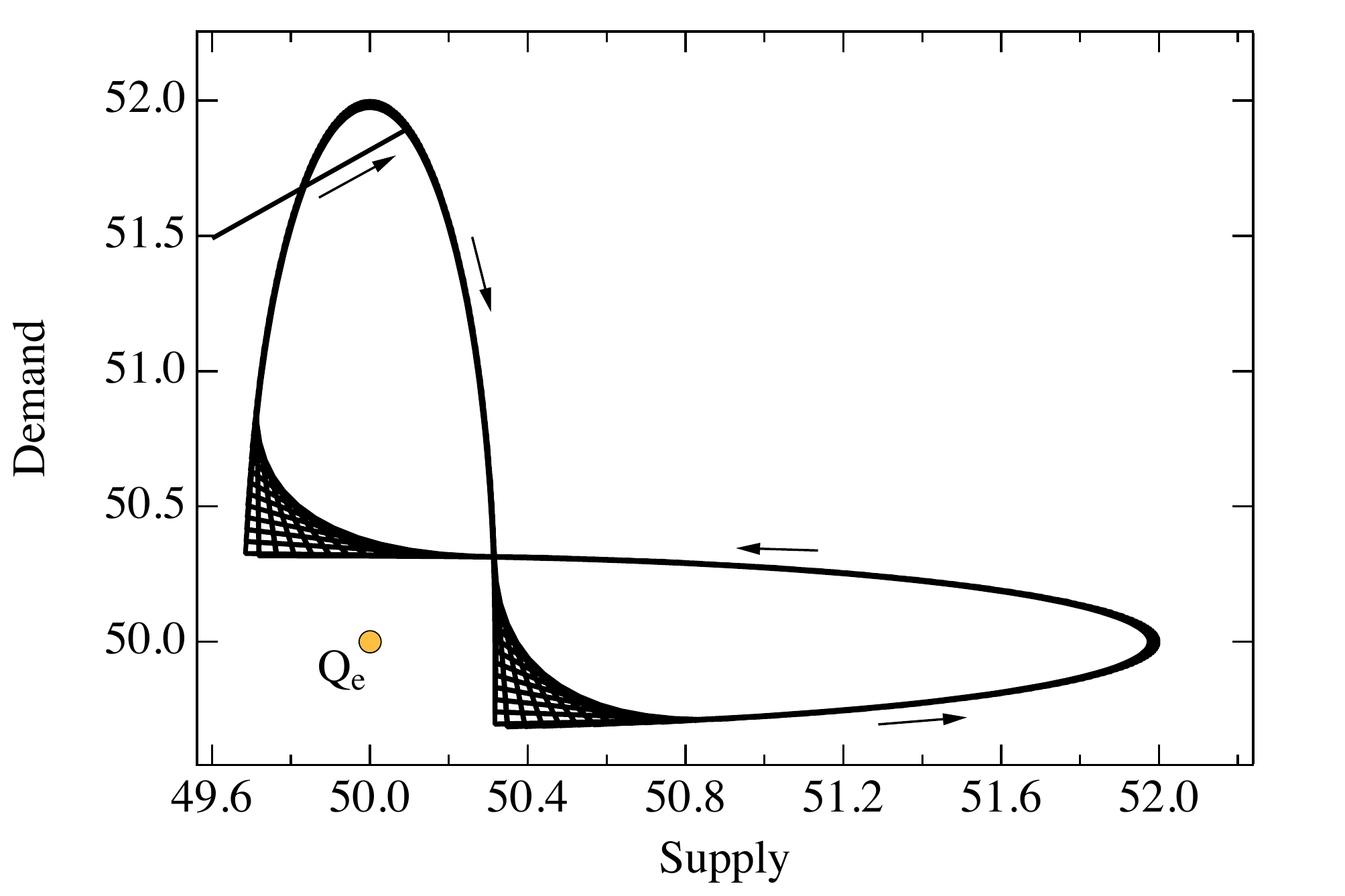}}
\end{array}
$
\caption{\textbf{Dynamics with speculators: $k_{sp}=1$} - Dynamic response of the system to a supply/demand shock at time $t=0$. \emph{Top Left}: Oscillations of $P$ around its equilibrium value. \emph{Top Right}: Supply and demand as a function of time. \emph{Bottom Left}: Price/quantity relationship. \emph{Bottom Right}: Dynamic evolution of $Q_d$ vs $Q_s$.}
\end{figure}

\noindent where $m_{1,2}=(-b\pm\delta^{1/2})/2$. If both $m_1$ and $m_2$ lie between 0 and 1 in absolute value, then both $m_1^t$ and $m_2^t$ approach zero, and the solution converges exponentially. Otherwise the solution exponentially diverges.\\

\textbf{Case 2: $\delta=0$}

If the discriminant is zero, then there is exactly one real root. The solution in this case is:

\begin{equation}
P(t)=(P_1-P_e)(-b/2)^{t-1}t+P_e
\end{equation}
Whether the behavior is convergent or divergent now depends just on the magnitude of $b$. However, the likelihood of the roots being exactly equal when dealing with economic data is extremely small. \\

\begin{figure}[t]
\refstepcounter{figref}\label{fig:spec_105}
\center
$
\begin{array}{cccc}
\href{http://necsi.edu/research/social/img/fig10a.pdf}{\includegraphics[width=0.5\linewidth]{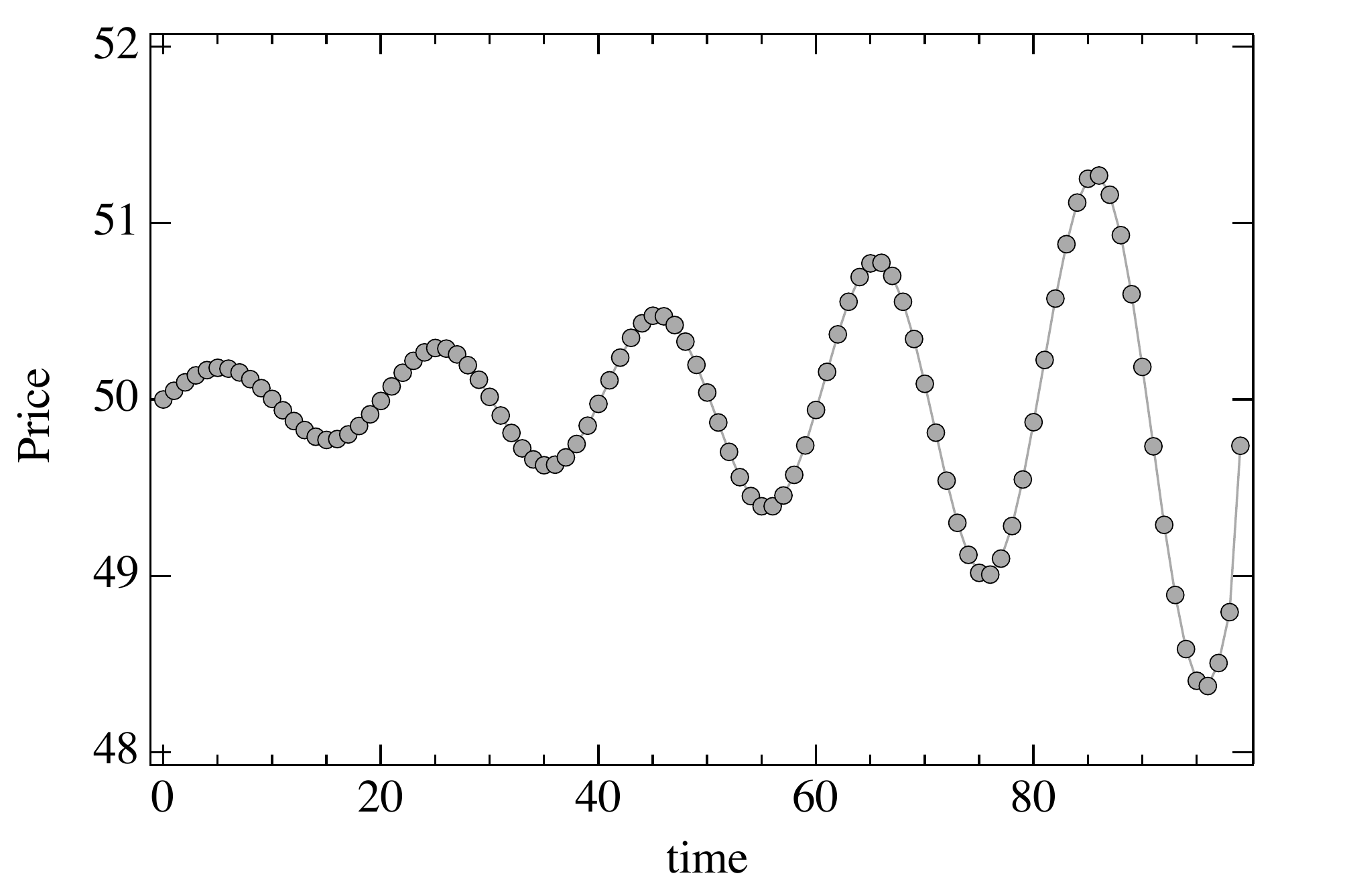}} &
\href{http://necsi.edu/research/social/img/fig10b.pdf}{\includegraphics[width=0.5\linewidth]{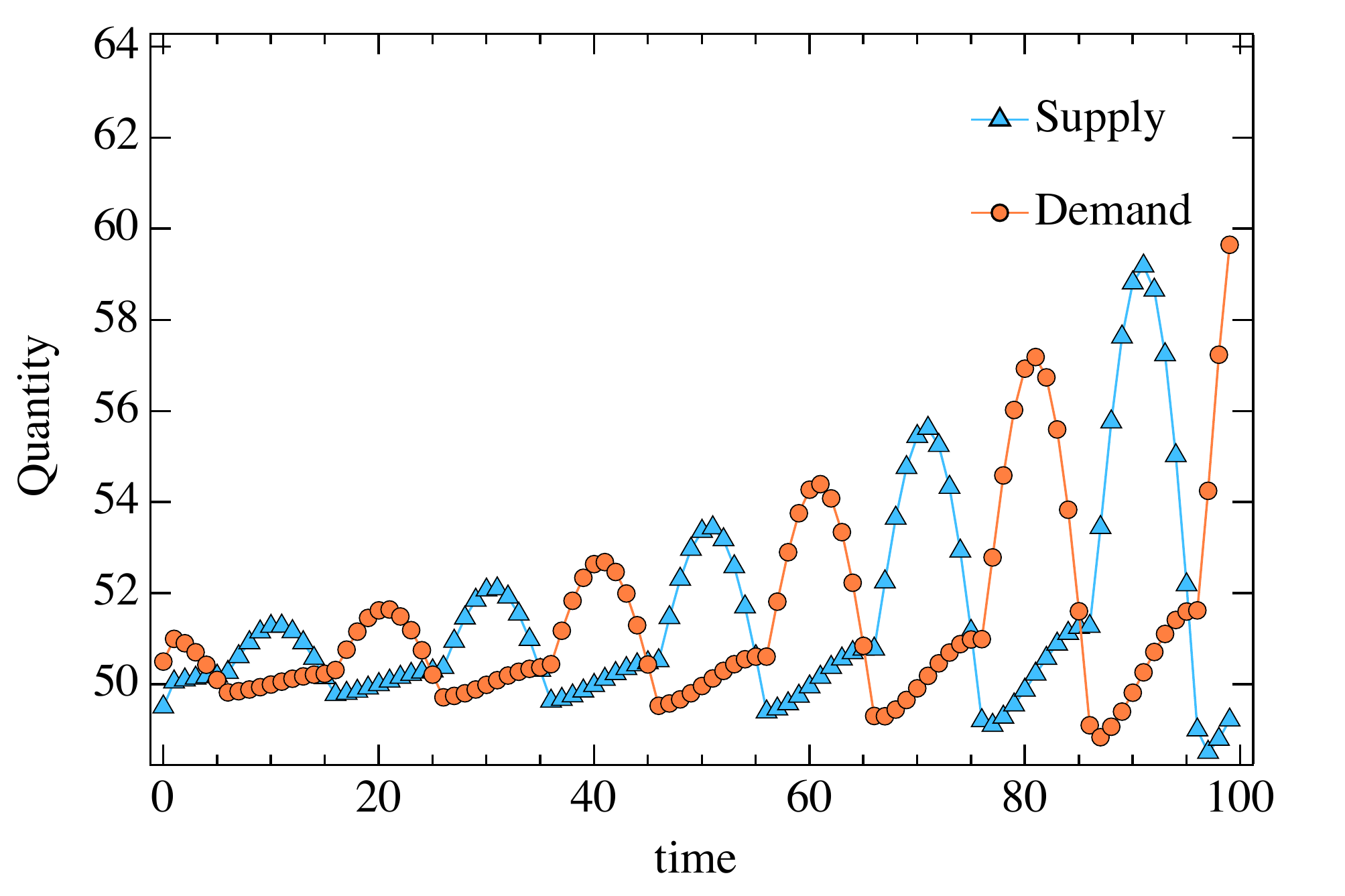}} \\
\href{http://necsi.edu/research/social/img/fig10c.pdf}{\includegraphics[width=0.5\linewidth]{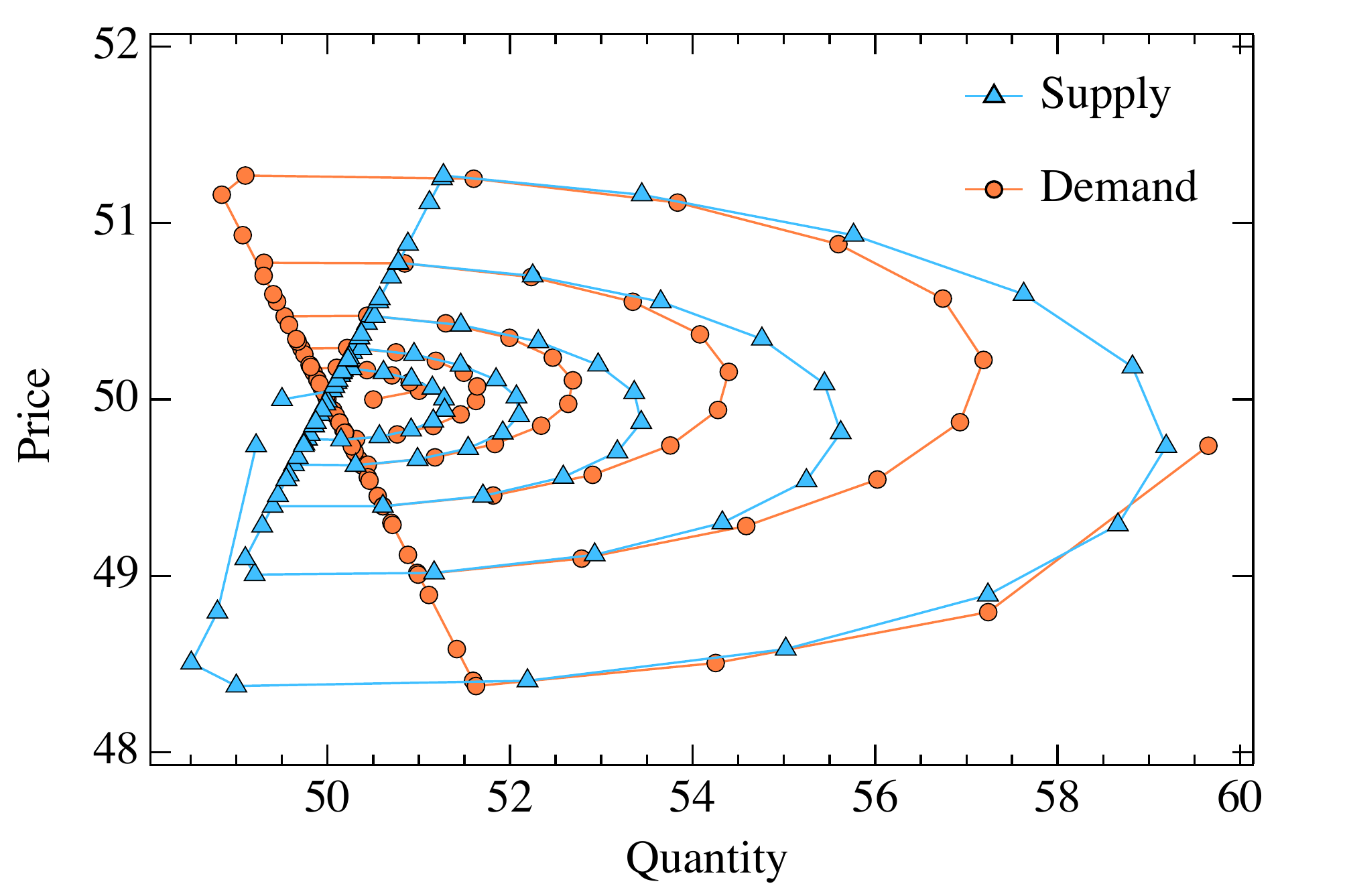}} &
\href{http://necsi.edu/research/social/img/fig10d.pdf}{\includegraphics[width=0.5\linewidth]{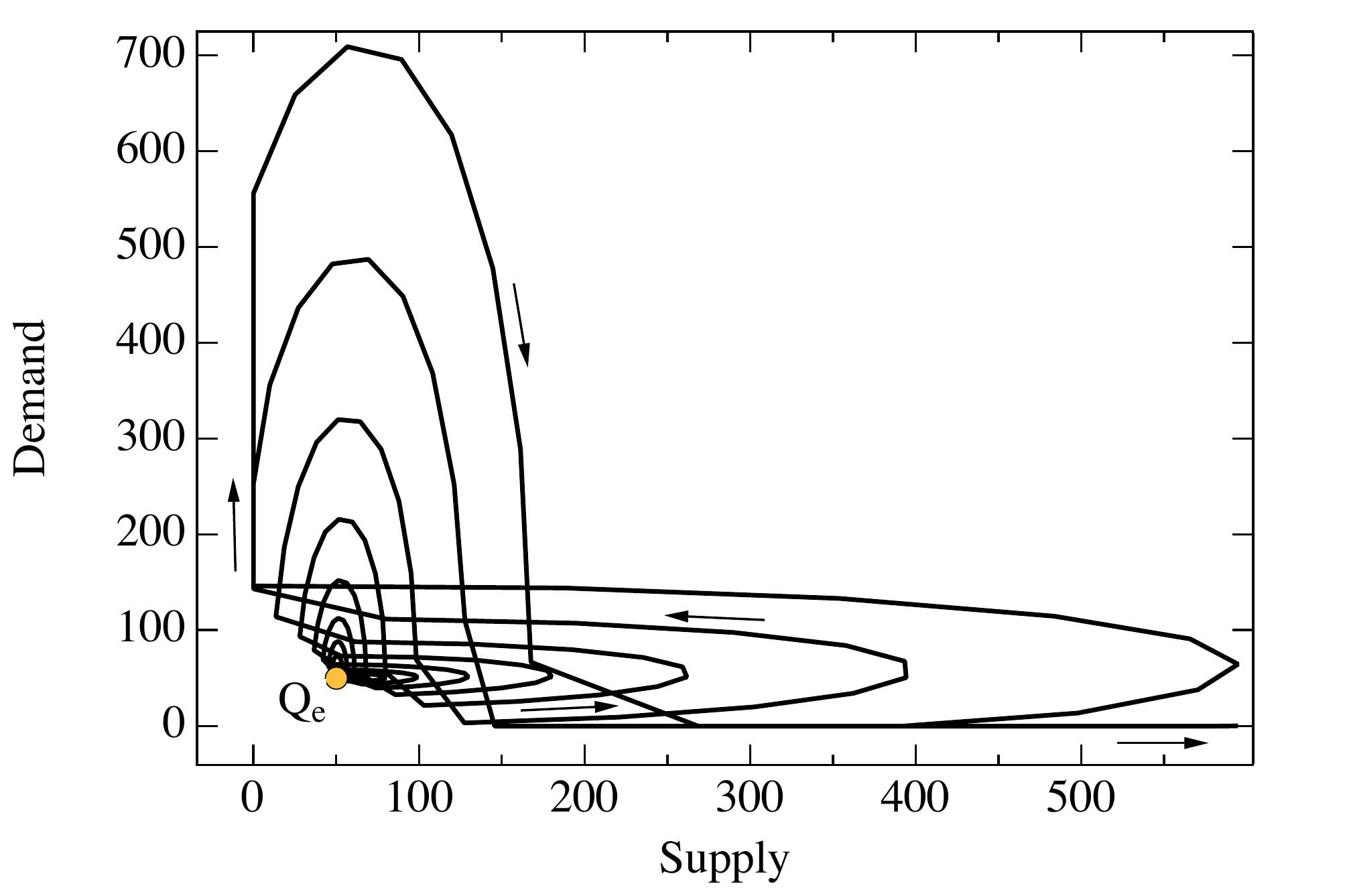}}
\end{array}
$
\caption{\textbf{Dynamics with speculators: $k_{sp}>1$} - Dynamic response of the system to a supply/demand shock at time $t=0$. \emph{Top Left}: Oscillating divergence of $P$ to its equilibrium value. \emph{Top Right}: Supply and demand as a function of time. \emph{Bottom Left}: Price/quantity relationship. \emph{Bottom Right}: Dynamic evolution of $Q_d$ vs $Q_s$ away from $Q_e$.}
\end{figure}

\textbf{Case 3: $\delta<0$}

If the discriminant is negative, the solution to Eq. \ref{eq:spec} becomes:

\begin{equation}
P(t)=(-\sgn (b))^{t-1}\sqrt{k_{sp}^{t}}\left(\frac{P_1-P_e}{\sqrt{k_{sp}}}\right)\frac{\sin(\theta t)}{\sin(\theta)}+P_e
\label{eq:case3}
\end{equation}
where
\begin{equation}
\displaystyle\theta=\arcsin{\sqrt{1-\frac{b^2}{4k_{sp}}}}
\label{eq:theta}
\end{equation}
and $\sgn()$ is the signum function. The behavior in this case is oscillating, with a period $T=2\pi/\theta$. Whether $P(t)$ converges to its equilibrium value (as in Fig. \ref{fig:spec_095}) or not (as in Fig. \ref{fig:spec_105}) depends on the growth factor $k_{sp}$. Given the necessary combinations of the four parameters of the supply-demand relationship ($\alpha_s,\alpha_d,\beta_s,\beta_d$), $k_{sp}$ remains the only relevant parameter. We can distinguish the price dynamics behaviors of the model according to the values $k_{sp}$ assumes when the discriminant is negative:

\begin{figure}[t]
\refstepcounter{figref}\label{fig:phase}
\href{http://necsi.edu/research/social/img/fig11.pdf}{\includegraphics[width=0.6\linewidth]{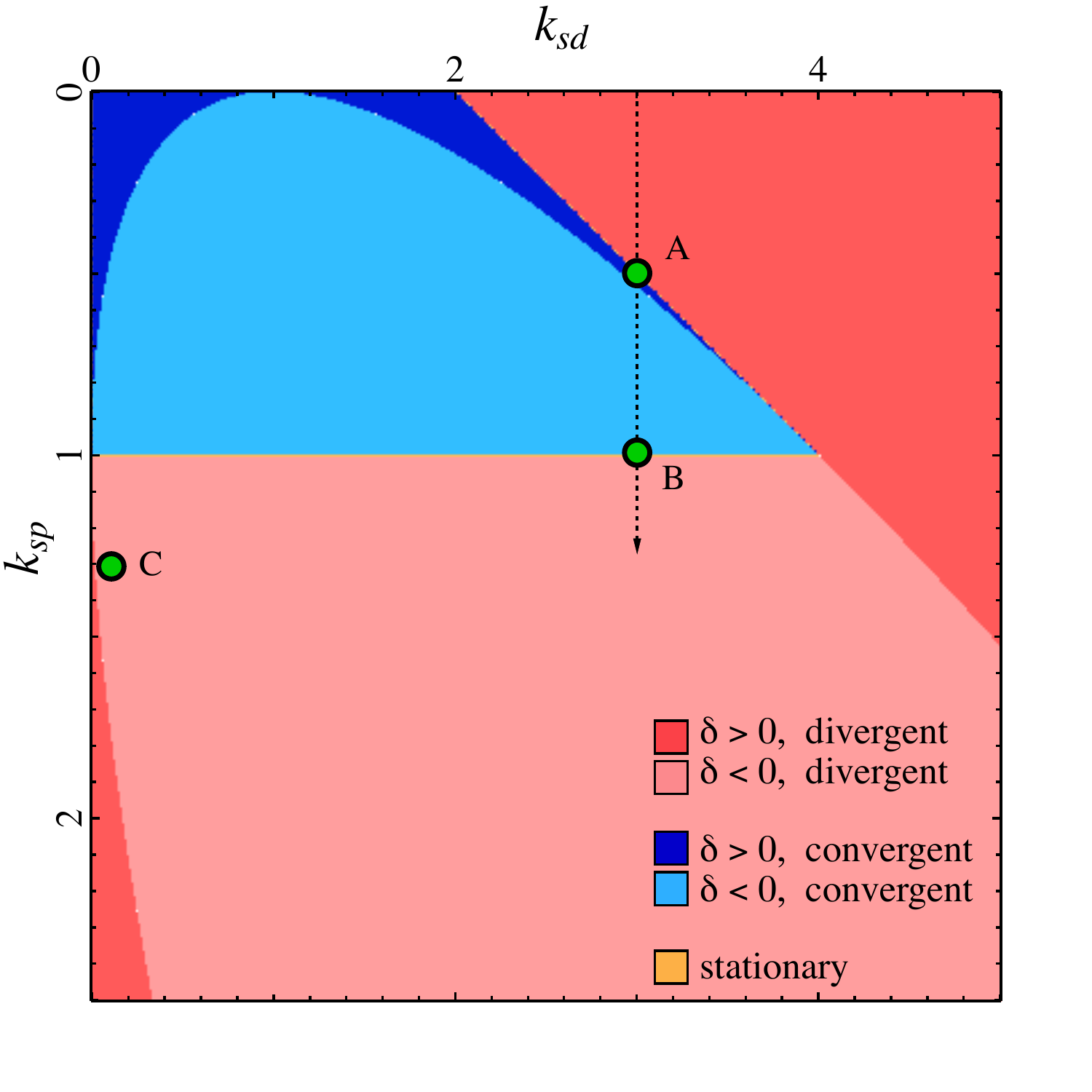}}
\caption{\textbf{Model Phase Diagram} - Behavior of the model for different values of its two main parameters, $k_{sd}$ and $k_{sp}$. Dark red regions correspond to a divergent behavior according to Eq. \ref{eq:case1}, light red to a divergent behavior according to Eq. \ref{eq:case3} (see also Fig. \ref{fig:spec_105}), light blue to a convergent behavior (Eq. \ref{eq:case3} and Fig. \ref{fig:spec_095}), as well as dark blue (Eq. \ref{eq:case1}). A thin yellow line between the light blue and light red regions defines a stationary point at $k_{sp}=1$ (Eq. \ref{eq:case3} and Fig. \ref{fig:spec_1}). The region from point A to point B represents the stabilizing effect of speculators as $k_{sp}$ increases at $k_{sd}=3$. C is the point in the phase space $(k_{sd},k_{sp})$ corresponding to the values obtained with the fitting of food price data (see Fig. \ref{fig:food_spec}).}
\end{figure}

\begin{tabbing}
$k_{sp}=0$ ~~~\=$P(t)$ decays exponentially to $P_e$ (Figure \ref{fig:nospec}) \\\\

$k_{sp}<1$ \>$P(t)$ converges to $P_e$ with damped oscillations (Figure \ref{fig:spec_095}) \\\\

$k_{sp}=1$ \>$P(t)$ oscillates around $P_e$ (Figure \ref{fig:spec_1})\\\\

$k_{sp}>1$ \>$P(t)$ diverges with amplified oscillations (Figure \ref{fig:spec_105})\\
\end{tabbing}

The behavior of the system is summarized in Fig. \ref{fig:phase}, where the phase diagram of the model is plotted as a function of its two main parameters: $k_{sd}$, the fundamental supply-demand contribution to price dynamics, and $k_{sp}$, the speculator contribution. The blue region on the top left corner is the stable region of the system, where price converges to its equilibrium value, while the red region around it defines the domain of price divergence.

The difference between the dark blue region and the light blue one is the sign of the discriminant. When $\delta$ is negative (light blue region), we have the damped sinusoidal behavior shown in Fig. \ref{fig:spec_095}; when $\delta$ is positive, we can either have the exponential decay shown in Fig. \ref{fig:nospec} (left-side dark blue triangle in the phase diagram) or a damped oscillating behavior (right-side dark blue triangle). The two triangles are separated on the $x$-axis ($k_{sp}=0$) by $k_{sd}=1$: in this case in fact, $\delta=(k_{sd}-1)^2$ and whether the behavior is oscillating or monotonic depends on the sign of the quantity in parentheses. On the other hand, if $k_{sd}>2$ the supply and demand elasticities are too high, $\delta>1$ and the price diverges (red region).

The question of whether speculators stabilize or destabilize prices has been the subject of a large body of literature \cite{ref:Hart_1986}, going back to Milton Friedman, who said ``People who argue that speculation is generally destabilizing seldom realize that this is largely equivalent to saying that speculators lose money, since speculation can be destabilizing in general only if speculators on average sell when the [commodity] is low in price and buy when it is high.'' \cite{ref:Friedman_1953}. Our simple model provides a quantitative assessment of the role of speculators: if we follow the arrow on the phase diagram from the $x$-axis at $k_{sd}=3$ and $k_{sp}=0$, for example, we see how increasing the effect of speculators may actually stabilize the system at first (from point A to point B), but eventually the system leaves the convergent behavior and becomes unstable again. Therefore a small amount of speculation may help prices to converge to their equilibrium value, but if the market power of speculators is too great they will have a destabilizing effect on the price dynamics. This holds true as long as the model parameter $k_{sd}<4$; otherwise speculators are never able to stabilize the market.

The condition for speculator induced instability of a supply and demand equilibrium, $k_{sp} \ge 1$, can be understood by recognizing that at $k_{sp} = 1$ the additional speculator activity motivated by a price change is precisely enough to cause the same price change in the next period of time. Such momentum of the price is quite reasonably the condition for speculator induced bubbles and crashes. Supply and demand restoring forces are then responsible for the extent of the oscillatory behavior. 

The concepts of equilibrium and trend following are manifest in trader strategies that are ``fundamental" and ``technical" \cite{LeBaron2002}. Fundamental investing relies upon a concept of target price, the expected value. Investors estimate the target price based on supply and demand and use it as a guide to buy or sell. Technical investing considers various patterns in the price time series, the primary of which is the trend of prices itself, which sets direction but not value, except in relation to that pattern. More generally, in a technical strategy, a shift by a constant amount of the price time series would not affect investor decisions to buy or sell. Our model maps these two types of investing behavior onto the first two possible terms in a series expansion of the equation for price change in terms of the prices at previous times. These two terms represent respectively the two different types of investing behavior. The first term has a price difference from a reference (the equilibrium price), and the second term has the difference of sequential prices in the past. The equilibrium price in the first term is the average over the expected target price of all fundamental traders. Even with, or rather because of, a large diversity of individual trader strategies, an aggregation over them can be expected to leave these two terms dominant. Aggregation incorporates the multiple tendencies of individuals, and the diversity across individuals. The aggregate over their decisions has these two primary price impacts. 

Finally, we consider the mechanisms by which trend following speculators are related to rational expectations about future prices and their impact on current prices and inventory. In the analysis of inventory changes over time in the ``supply of storage'' model \cite{sos_1,sos_2}, it has been shown that inventories increase when future prices are expected to rise. The inventory change is then achieved by a departure of prices from supply and demand equilibrium at that time. However, this is due to a \emph{future} supply and demand change. In effect this analysis is the basis of all trading that achieves price stability over time due to inventory. Thus, if there is a seasonal supply of grain, the storage of that grain for future use is motivated by a difference in the timing of demand, and prices are adjusted to the demand across time. Given a temporary expected higher demand or lower supply at a time in the future, prices may be adjusted at the current time to sell less grain in order to keep the grain for the future. 

Trend following reflects the assumption, as indicated in the speculator model, that extrapolation is a valid representation of expected future prices (including the possibility that the trends represent actual changes in supply and demand). Under these conditions it is rational to increase prices in order to reserve inventory for the future prices, causing a departure from equilibrium. This increase in price caused by the expected future prices then leads to a more rapid increase in price. Absent a way to distinguish the increase in price that is due to the desire to adjust inventory from other increases in price, we now have a recursive process. This is exactly the problem of recursive logic leading to multiple possible truths or self-contradicting paradox. Interpreted as a dynamical system, because of iterative rather than synchronous steps, the result is the dynamics of bubble and crash behavior described above. In particular, the trend following trader assumption of extrapolated trends predicting future price increases is inherently (globally) irrational due to its recursive tendency toward infinite or zero prices, only moderated by supply and demand traders. This does not imply that it is not locally rational, i.e. contextually or at a particular time it is a rational behavior, but any attempt to generalize local to global rationality encounters analytic problems. The absence of rationality is manifest \emph{a posteriori} in empirical data by the occurrence of crashes after bubbles.  Our analysis, however, shows that an empirical crash is not necessary to prove irrationality of trend following because of its inherent paradoxical nature. Nevertheless, as we found in our model, a limited amount of trend following can improve market behavior, in essence because trend following has a limited degree of validity in rational prediction of future prices. We might say that a small amount of an irrational behavior can contribute to increased rational collective action.

As discussed in Appendix \hyperref[app:e]{E}, using food price data, we find the current world market to be at point C in the phase diagram. This is a region where the price diverges with amplified oscillations. In this domain, speculation can strongly destabilize the supply and demand equilibrium price.


\newpage
\begin{center}
\phantomsection
\label{app:e}
\Large Appendix E\\
\Large Food Price Model: Speculators and Ethanol Demand
\end{center}

We construct an explicit model of price dynamics to compare to the food price index. Since our analysis has eliminated all supply and demand factors except ethanol conversion as a major shock, and the only other factor of known relevance is speculators, our model is constructed in order to represent these two effects. We build the simplest possible model of these two factors, minimizing the number of empirically adjustable parameters, and find a remarkably good fit between theory and empirical data. 

We combine the ethanol model described in Appendix \hyperref[app:c]{C} and the speculator model described in Appendix \hyperref[app:d]{D}. We consider only the FAO food price index to characterize the combined effect on food prices. Because 
a majority of financial holdings in agricultural futures markets are now due to commodity index funds \cite{Worthy2011}, 
it is reasonable to model aggregate effects of speculators on commodities rather than on individual ones separately. Similarly, corn ethanol conversion impacts food prices through a number of parallel mechanisms. The mutual influences of grain prices through substitution and replacement, as well as geographical heterogeneity of individual countries or regions, require detailed modeling that need not be done at a first level of representation. 

Starting from the supply and demand model with Walrasian adjustment
\begin{equation}
P(t+1)=k_c+[1-k_{sd}]P(t)
\end{equation}
we include the effects of assuming a dominant ethanol conversion demand shock from Appendix \hyperref[app:c]{C}. Since the equilibrium price is given by $P_e=k_c/k_{sd}$, we constrain $k_c$ to be:
\begin{equation}
k_c(t)=(a+bt^2)k_{sd}+b(2t+1)
\end{equation} 
where $a$ and $b$ are the coefficients of the corn ethanol model obtained in Appendix \hyperref[app:c]{C}. The factor $(a+bt^2)$ is the time-dependent equilibrium price from the corn ethanol model. The additional term $b(2t+1)$ corrects for the lag in update of the dynamic model with respect to the equilibrium model, causing the dynamic model to track the equilibrium model price rather than a price that is lower, i.e. lagging in time, during the initial period. This term does not substantially affect the overall fit of the speculator and ethanol model. 

\begin{figure}[b]  
\refstepcounter{figref}\label{fig:food_spec}
\href{http://necsi.edu/research/social/img/fig12b.pdf}{\includegraphics[width=0.735\linewidth, trim= -1.49cm 1.8cm 0cm 0.3cm, clip]{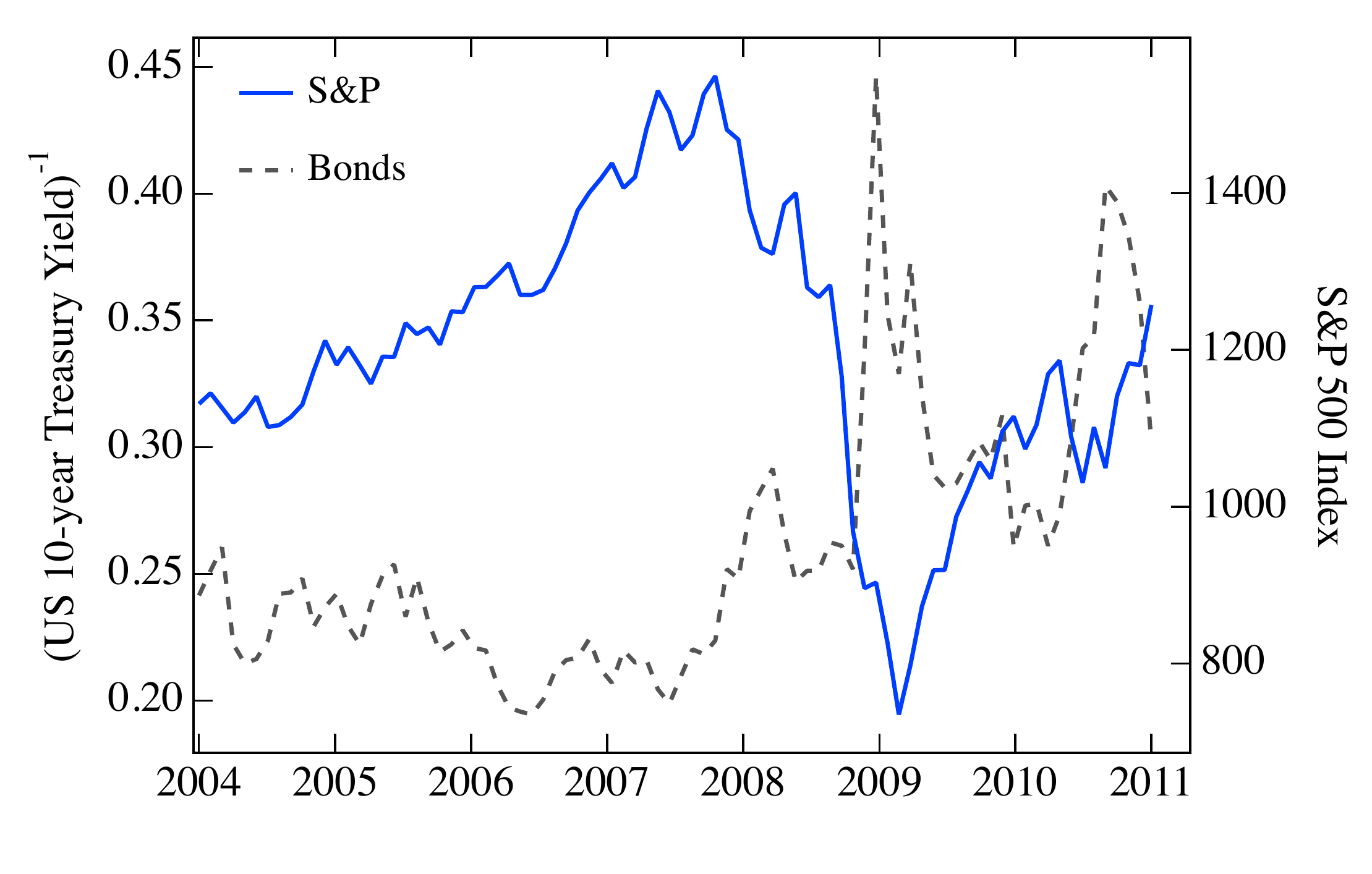}}\\
\href{http://necsi.edu/research/social/img/fig12a.pdf}{\includegraphics[width=0.6\linewidth, trim= 0cm 0cm 0cm 0cm, clip]{figs/food_spec.pdf}}
\caption{\textbf{Model Results} - \emph{Top}: Solid line is the monthly FAO Food Price Index between Jan 2004 and Apr 2011. Dashed line is the best fit according to Eq. \ref{eq:model}. The effect of speculators is turned on in the first half of 2007, when the housing bubble collapsed. Values of parameters are: $k_{sd} = 0.098$, $k_{sp} = 1.29$, $\mu_{equity}\gamma_0 = - 0.095$, $\mu_{bonds}\gamma_0 = -67.9$. \emph{Bottom}: Time series used as input to the speculator model, S\&P500 index (stock market, in blue) and inverse of the US 10-year Treasury Note Yield (bonds, in black).}
\end{figure}

We incorporate the effect of trend following speculators as in Appendix \hyperref[app:d]{D} by adding a term $k_{sp}[P(t)-P(t-1)]$ which interacts with the price dynamics due to the supply and demand terms. 
The last step in our construction of the speculator model is to add the effect of alternative investment markets on the price of the commodity. We assume that when the price change of an alternative investment is  positive in the previous time step, speculators sell a quantity $\mu_i[P_i(t)-P_i(t-1)]$ of commodity contracts, where $P_i(t)$ is the price at time $t$ of investment $i$ and $\mu_i<0$, in order to shift part of their capital to the new market. This sale of commodities competes against the purchase of commodities given by $\mu_i[P_i(t-1)-P_i(t)]$, representing the maximum profit seeking behavior of speculators who transfer capital between markets. In summary, Eq. \ref{eq:spec} becomes: 
\begin{equation}
P(t+1)=k_c(t)+[1-k_{sd}]P(t)+k_{sp}[P(t)-P(t-1)]+\sum^{N}_{i=1}k_i[P_i(t)-P_i(t-1)]
\label{eq:model}
\end{equation}
where $N$ is the number of alternative investments taken into account, and $k_i=\mu_i\gamma_0$ are the alternative investment coupling constants. The model has effectively $N+3$ fitting parameters: two deriving from supply and demand considerations ($k_c$ and $k_{sd}$) and $N+1$ deriving from trend-following considerations. 

In Fig. \ref{fig:food_spec}, we show the best fit of this model to the FAO Food Price Index. We start the fit in 2007, when speculators presumably started moving their investments from the stock market to other markets, as suggested by the bubble dynamics of Fig. \ref{fig:bubbleprotests}. 
The date of the start, May 2007, is chosen for best fit. A more gradual increase in investor interest would be more realistic and represent the data more closely, but the simple model using a single date for investor interest is sufficient. 
The alternative markets we consider besides commodities are equities (using the S\&P500 Index time series) and bonds (using the US 10-year treasury note price time series), which have peaks right before and right after the peak in the commodity time series (top panel of Fig. \ref{fig:food_spec}) so that $N=2$. 

The resulting price curve is constructed directly from the model using only the adjustment of four model parameters ($k_{sd}$, $k_{sp}$, and the two market coupling parameters, $k_1$  and $k_2$) and the alternative market prices as input. The two large peaks are precisely fit by the model, as is the intermediate valley and smaller intermediate peak. The stock market plays a key role in the fit due to a shift of investment capital in 2009 in response to a stock market increase. The bond market plays a smaller role and the coefficient of coupling between the commodity and bond markets is small. The parameters that are obtained from the fitting $(k_{sd},k_{sp})$ are shown as point C in Fig. \ref{fig:phase}. The point lies in the unstable region of the system, with the caveat that we fit the Food Price Index with Eq. \ref{eq:model} that includes the alternative markets but we plot the phase diagram for Eq. \ref{eq:spec} without those markets.

Our results for the Food Price Index yield parameters that can be compared with expectations about speculator influence on commodity markets. In particular, the value of $k_{sp} = 1.29$ is consistent with a speculator volume that can move prices 30\% more than the price change that is found in the previous time.  

The value of the supply and demand parameter $k_{sd}=0.098$ combines with the speculator behavior to yield a bubble and crash cycle of $2\pi/\theta = 23.6$ months (see Eq. \ref{eq:theta}), almost exactly two years, consistent with a single year of price increases. This corresponds to the natural assumption of an annual cycle for the maturation of futures contracts for delivery that impact on actual supply and demand, a financial planning time of a year \cite{Wright2011, Hostetter2011}.
The maturation of such contracts leads to increases in inventories
and thus a restoring force toward supply and demand equilibrium.
Furthermore, according to this analysis, we predict an increase in inventories
of grains starting at the peak of the speculative bubble, one year after the departure from equilibrium prices.  As shown in Fig. \ref{fig:stocks}, this is consistent with the available data on observed inventories
of grains \cite{source_grains}. In particular, the inventories
increased from September 2008 to September 2009.  

The result that our dynamic speculator model is able to fit the FAO Food Price Index and that the supply-demand model of Appendix \hyperref[app:b]{B} is not able to do so is consistent with the hypothesis that speculators played an important role in determining food prices.
In conjunction with the other evidence for speculator involvement (see main text), our quantitative model provides specific evidence not just for a role of speculators, but for the extent of impact of speculators on the food and other commodity markets.

\bibliographystyle{Science}

\end{document}